# Investigation of cation self-diffusion mechanisms in UO$_{2\pm x}$ using molecular dynamics

A.S. Boyarchenkov [a], S.I. Potashnikov [a], K.A. Nekrasov [a], A.Ya. Kupryazhkin [a]

[a] Ural Federal University, 620002, Mira street 19, Yekaterinburg, Russia

potashnikov@gmail.com  boyarchenkov@gmail.com  kirillnkr@mail.ru  kupr@dpt.ustu.ru

**Abstract**

This article is devoted to investigation of cation self-diffusion mechanisms, taking place in UO$_2$, UO$_{2+x}$, and UO$_{2-x}$ crystals simulated under periodic (PBC) and isolated (IBC) boundary conditions using the method of molecular dynamics in the approximation of rigid ions and pair interactions. It is shown that under PBC the cations diffuse via an exchange mechanism (with the formation of Frenkel defects) with activation energy of 15–22 eV, while under IBC there is competition between the exchange and vacancy (via Schottky defects) diffusion mechanisms, which give the effective activation energy of 11–13 eV near the melting temperature of the simulated UO$_{2.00}$ nanocrystals. Vacancy diffusion with lower activation energy of 6–7 eV was dominant in the non-stoichiometric crystals UO$_{2.10}$, UO$_{2.15}$ and UO$_{1.85}$. Observations showed that a cation vacancy is accompanied by different number of anion vacancies depending on the deviation from stoichiometry: no vacancies in UO$_{2.15}$, single vacancy in UO$_{2.00}$ and four vacancies in UO$_{1.85}$. The corresponding law of mass action formulas derived within the Lidiard-Matzke model allowed explaining the obtained activation energies and predicting a change in the activation energy within the temperature range of the superionic phase transition. The diffusion of cations on the surface of nanocrystals had activation energy of 3.1–3.6 eV.

**PACS number(s)**: 28.52.Fa, 66.30.-h, 61.72.J-, 64.70.kp, 83.10.Rs

**Keywords**: molecular dynamics, pair potentials, uranium dioxide, free surface, uranium self-diffusion, activation energy, Frenkel defect, Schottky defect, superionic transition.

## *1. Introduction*

In ceramic and ionic materials diffusion of cations and anions occurs in separate sublattices [1], because positions in cationic sublattice are energetically unfavorable for anions and vice versa.

Self-diffusion of oxygen anions in uranium dioxide is a relatively fast process, so it has been well studied both experimentally (see reviews [2] [3]) and by computational modeling (see our review [4] and the works of other authors [5] [6] [7]). However, the phenomena associated with the mass transport (grain growth, sintering, creep, plastic deformation, recrystallization, etc.) include movement of ions of both types, so their rate is determined by much slower cation diffusion [8]. This relationship is confirmed experimentally for creep, grain growth and sintering in oxides, carbides, nitrides and carbonitrides (see [2] [9]).

Such phenomena can lead to deviation from the optimal operation mode, change of segregation rate of radioactive fission products (RFP) and accumulation of defects, thereby impairing mechanical and electronic properties of the fuel. In addition, fuel burnup significantly alters transport properties and thermal conductivity [10]. Therefore, in order to improve safety of nuclear reactors and minimize accident-related environmental damage it is important to study the diffusion of cations over a wide range of temperatures from 700–1500 K in the operating mode up to the melting point of 3150 K (due to Reactivity-Initiated Accidents, e.g. after the Loss Of Coolant).

Diffusion of cations in UO$_2$ and related compounds (PuO$_2$, ThO$_2$, MOX) is very slow (less than $10^{-15}$ cm$^2$/s [2] or $10^{-17}$ cm$^2$/s [11]) even at high temperatures of 1800–2000 K, which are among the maximum temperatures reached in the relevant experiments corresponding to the crystalline phase. Superionic phase transition, during which anionic sublattice become completely disordered, occurs at higher temperatures of 2600–2700 K (in UO$_2$), where there are no direct experimental data on diffusion. We only know that the abnormal growth of creep rate near the superionic transition can not be explained by the anionic sublattice disordering or linear extrapolation of the low-temperature data [12]. Thus, the question of the impact of the superionic phase transition on the cationic subsystem remains open.

In the existing "low-temperature" experiments (see [2] [11]) a variety of methods were used, including direct (using radioactivity) and indirect (using data on the kinetics of sintering) approaches. At that, the published results have significant variation (three orders of magnitude of the uranium diffusion coefficient counting results for single crystals only). The higher volume diffusion coefficients previously determined in polycrystalline

UO$_2$ correspond to the uranium grain boundary diffusion due to some short-circuiting mechanism, as shown by Sabioni et al. [13]. According to Matzke [2] (see also [3]), the main sources of errors in the experimental measurements are: high rate of evaporation, formation of striated surfaces due to anisotropy of the surface energy, too thick layers of tracer, deviation from stoichiometry, the use of sintered material with small grain size, etc.

Computational modeling of diffusion using the method of molecular dynamics (MD), which allows investigating ion transport under precisely known and controlled conditions, is not affected by the complexities of nature experiments, so it may help to clarify characteristics of uranium cations movement in the real crystals. However, due to the limitations of computational tools, so far MD-simulations of crystals without artificial defects did not so far allowed to register the intrinsic disordering of the uranium sublattice. Therefore, it was considered since the earliest works on MD-simulation of UO$_2$ [14] [15] [16] that the non-zero slope of the mean square displacement (MSD) of uranium ions *vs.* time indicates the molten state of the system.

Our previous works [17] [4] showed that the experimental data on anion diffusion are best reproduced by simulation of nanocrystals isolated in vacuum. In 2008 the first attempt of calculation of cation diffusion coefficient (DC) in such nanocrystals was made [18]. The diffusion coefficients obtained for cations in the bulk were about $10^{-9}$–$10^{-7}$ cm$^2$/s and seemed independent of interatomic potentials (Walker-81 and Nekrasov-08) and of system size (4116 and 6144 ions). However, the simulation time in that work was limited to ~100 ps, which was too small for registering vacancy diffusion of cations. Besides, UO$_2$ nanocrystals of the cubic shape used there have considerable excess energy and, as shown in [19], such crystals are subject to a process of structural relaxation. This process does not affect the faster diffusion of anions (see the comparison of diffusion in the cubic and octahedral crystals [4]), however, led to a significant overestimation of cation DCs obtained in [18].

In [20] the uranium diffusion in the presence of grain boundaries was simulated. Cation diffusion hops were registered only at a distance of less than 12 Å (approximately two lattice periods) from the grain boundary. The diffusion coefficients were in the range of $10^{-9}$–$10^{-7}$ cm$^2$/s, and the diffusion activation energy was only 0.8 eV. This energy is significantly lower than all the theoretical estimates of the bulk cation migration enthalpies (see, for example, [21] [22]) and 4–6 times smaller than the diffusion activation energy measured in experiments with polycrystals [23] [13].

In [5] the uranium diffusion was simulated both in a system with embedded Schottky defects and in a system with grain boundaries. In the first case, the DCs were in the range $10^{-11}$–$10^{-9}$ cm$^2$/s, and the authors stated that extrapolation of the results to lower temperatures give values comparable with the experimental data. However, the corresponding activation energy was not provided (only its estimate by lattice statics method was given). In the case of grain boundaries, the activation energy of 2–4 eV and much higher DCs were obtained. These latter results were "almost comparable" to the experiments of Reynolds [24] and density functional theory (DFT) calculations. The authors also noted that the behavior of both types of ions at the grain boundaries were similar to behavior in the liquid phase.

Finally, in [25] migration of cavity in UO$_2$ was simulated in the presence of temperature gradient. Previously three possible mechanisms were proposed for cavity migration: via surface diffusion, via vacancy diffusion in the crystal lattice, and via evaporation-condensation. Experimental studies did not allow identifying the dominant mechanism, whereas the simulation indicated a surface mechanism. The cation diffusion activation energy of 2.66±0.05 eV was obtained, although atoms that moved by more than 1.5 of the lattice constant were taken into account only. Evaporation was not found, and the estimate of the lattice diffusion in the presence of randomly distributed cation vacancies was $3\times10^{-12}$ cm$^2$/s at a temperature of 2900 K.

The aim of this work is to:

- obtain reliable cation diffusion coefficients for the uranium dioxide system without artificial defects (i.e. diffusion via intrinsic defect formation), using high-performance graphics processors and the original methodology, successfully tested on the diffusion of anions [17] [4];
- identify and characterize the cation diffusion mechanisms occurring in the model crystals using visual observations and analysis of diffusing ions trajectory analysis;
- study the effect of boundary conditions on the diffusion of cations, expecting that PBC will allow examining of the exchange diffusion mechanism, and that under IBC Schottky defects participating in the uranium vacancy mechanism will be formed on a free surface of the model nanocrystals;
- estimate the effect of temperature-independent (not intrinsic) defects by comparing cation diffusion mechanisms in UO$_2$, UO$_{2+x}$ and UO$_{2-x}$;

- assess the impact of the superionic transition on the activation energy of diffusion of cations.

## 2. Methodology

As in our previous studies [26] [27] [4], molecular dynamics simulations were carried out in the approximation of rigid ions. However, due to the high computational complexity of the task only five of the ten sets of pair potentials (SPPs) were chosen: Goel-08 [28], fitted in the harmonic approximation to the elastic properties at zero temperature; Morelon-03 [29], fitted using the method of lattice statics to the formation energy of point defects; Basak-03 [30], MOX-07 [31] and Yakub-09 [32], fitted via MD simulation to the thermal expansion and the bulk modulus. See parameters of all these potentials in the first article [26]. These potentials provide the lowest melting point and the lowest energy of formation of cation Frenkel defect (since it correlates with the exchange diffusion activation energy and thus determines the slope of DC temperature dependence). In addition, four of these five SPPs reproduce the thermophysical properties of $UO_2$ better than the rest [26].

In this study we also used the technology of parallel computing on high-performance GPU [17] [33] [34] [35]. This approach allowed us to calculate the first reliable temperature dependences of cation DC in the absence of artificial defects, which required very long (on the scale of nanoseconds and microseconds) simulation time. These temperature dependences were calculated with step of 1–10 K, necessary to monitor the temperature dependence of the diffusion activation energy and changes in migration mechanism.

In order to integrate Newton's equations of motion, the Verlet method (with time step of 5 fs) and the Berendsen thermostat (with a relaxation time of 1 ps) were used. In simulations under PBC the volume of the system was controlled by the Berendsen barostat. Under IBC the system volume was not fixed, so nanocrystal simulations were carried out under NPT ensemble with zero ambient pressure.

### 2.1. Calculation of point defects formation energy

In order to calculate the point defects formation energies by the method of lattice statics, a gradient optimization of ion coordinates under PBC was used. In contrast to the Mott-Littleton approach (see, e.g., [37]), the relaxation was carried out throughout the whole crystal lattice (instead of just inner sphere) so that the calculated values correspond to intrinsic defects which could arise during the MD simulation.

The defects studied by the lattice statics were: the Frenkel defect (FD – interstitial cation and cationic vacancy), the anti-Frenkel defect (AFD – interstitial anion and anionic vacancy) and the Schottky defect (SD – electroneutral combination of one cationic and two anionic vacancies). Each energy value was calculated in stoichiometric and electroneutral crystal, where the interstitials and the vacancies were formed at the maximum distance possible within periodically-translated supercell. In the works of other authors (e.g., [38] [39] [37]) another method of calculation was usually used: in the case of Frenkel defect, for example, energies of the crystal with vacancy and the crystal with an interstitial ion were calculated independently to prevent interaction of the two defects, and their sum was taken as the FD formation energy.

In order to eliminate the influence of defect "reflections" appearing under PBC, we calculated the dependences of the energies on the supercell size, which ranged from 4×4×4 up to 16×16×16 unit cells (the number of unit cells per edge of the supercell is denoted by C below). These dependences were found to be linear with respect to 1/C (see [4]), which allowed us to calculate the defect formation energies at infinite separation distance (between the vacancy and the interstitial ion in case of either FD or AFD and between the vacancies in case of SD). The similar methodology of defect energy calculation was used in the recent work of Devynck et al. [40].

### 2.2. Calculation of the diffusion coefficients of volume and surface ions

The diffusion coefficient D of intrinsic ions at a given temperature $T$ was calculated from linear time ($t$) dependence of the mean square displacement (MSD) of ions by the Einstein relation:

$$\lim_{t \to \infty} \langle \Delta R^2 \rangle = 6tD(T) + const$$
$$\langle \Delta R^2 \rangle = \frac{1}{N} \sum_i \left| R_i^t - R_i^0 \right|^2 \quad (1)$$

The initial positions $R^0$ were updated every few nanoseconds in order to eliminate systematic errors, caused by arbitrary choice of starting moment of time and computational errors accumulation as well.

On the assumption of a persistent diffusion mechanism, the temperature dependence of the diffusion coefficient is expressed by the Arrhenius equation $D(T) = D_0 \exp(-E_D / kT)$, where the diffusion activation energy $E_D$ and the pre-exponential factor $D_0$ are constant characteristics of the process. In practice this equation holds approximately for the temperature intervals where a distinguished mechanism of migration exists.

Calculations of cation DC were firstly carried out under PBC (the model of quasi-infinite crystal without a surface), which were also used in most of the works of other authors. Then we moved to the model of nanocrystals isolated in vacuum (IBC), which had a free surface of energy-optimal octahedral shape [19], in order to study vacancy diffusion and the effect of boundary conditions. We will begin discussion of the DC calculation technique from the points relevant to both isolated and periodic boundary conditions.

The rate of diffusion jumps decreases exponentially with decreasing temperature. This fact requires exponential increase of either simulation time or supercell size in order to preserve statistically acceptable accuracy of DC measurement via MSD. In the case of too rare jumps linear MSD curves become stepped [4], and in the absence of diffusion jumps the MSD values characterize only thermal oscillations instead of diffusion. Hence, a correct and effective calculation of DC requires controlling the count of diffusion hops, e.g. via the MSD quantity [4].

The influence of system size was studied by variation of the ion count from 324 up to 12 000 particles. Similar to the previous simulation of $UO_2$ thermophysical properties [26], the diffusion coefficients of the system of 1500 particles under PBC were quantitatively almost the same as DCs of larger systems, and for half of the ten SPPs considered the differences between the systems of 768 and 1500 particles were quite small too (see the chart for anion diffusion in [4]).

Under IBC effect of the system size should be studied not only for the two types of ions, but also for the two regions (surface and bulk). This task is complicated by the lower melting point of isolated nanocrystals [27], because at lower temperatures the DC calculation takes a lot longer. The calculation of DC temperature dependence of anions showed that at temperatures above 1500 K the values for crystals of 768 and 12 000 particles differed by a factor of 1.5–2 only (when using MOX-07 SPP). Fig. 1 shows that the surface diffusion of uranium is much more dependent on the system size, due to the different mobility of ions on vertices, edges and faces of the crystal. Unfortunately, for the bulk diffusion of uranium similar exhaustive calculations are too consuming, therefore we give in Fig. 2 only a comparison of MSD evolution at one temperature of 3030 K, which is close to the melting point of the octahedral nanocrystal of 4116 particles. In general, one can say that for the systems of over 4116 ions the differences in DC on the chart with a logarithmic scale are negligibly small.

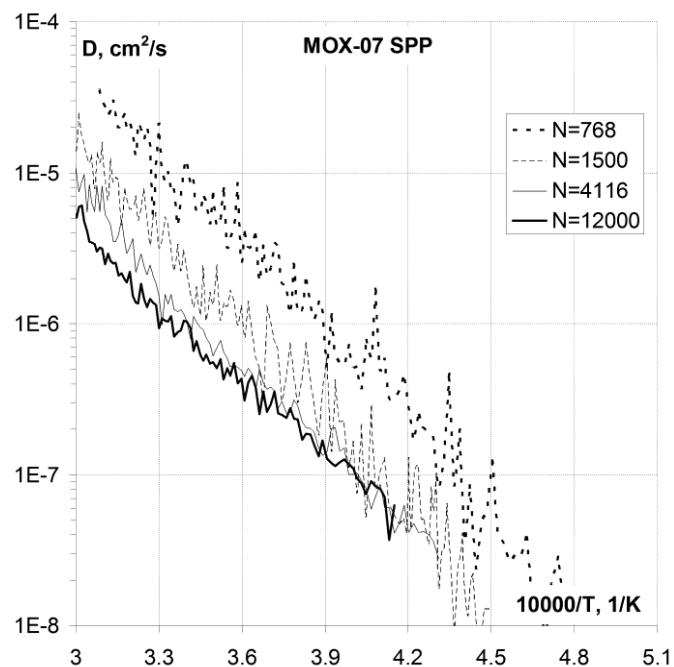

Fig. 1. Influence of nanocrystal size on temperature dependence of surface cation self-diffusion coefficient.

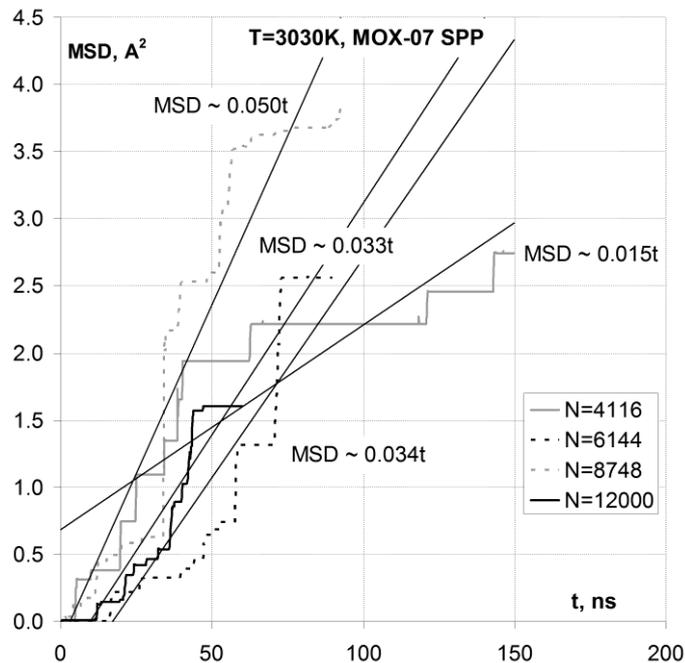

Fig. 2. Influence of nanocrystal size on time evolution of MSD of volume cations.

Thus, in order to accumulate statistics under IBC as well as under PBC it is more profitable to increase the simulation time instead of the system size, because the workload depends linearly on the number of MD-steps, but quadratically on the number of particles.

Now, let us consider unusual features of DC calculation that are specific to crystals with free surface simulated under IBC.

Particles at the crystal surface miss some neighbors, so their mobility (i.e., self-diffusion coefficient) differs from atoms in the bulk. Assuming that the mobility of a particle depends on its distance to the surface, it seems logical to divide crystal into layers (of cubic shape for cubic crystal, of octahedral shape for octahedral crystals, of spherical shape for the melt), so that particles of one layer will have approximately the same distance to the surface. Then, DC calculated for one individual layer will reflect the movement of each particle of this layer. Nevertheless, correct calculation of DC also requires taking into account some other factors.

The most important factor is that the crystal shape should be equilibrium [19]. For example, our experiments have shown that non-equilibrium cubic crystals (as in [18]) loose the original shape after 100 ps (at 2000 K) from the start; at that, not only surface atoms rearrange, but also the deeper layers. As a result, the original layering of the crystal becomes invalid, and the calculated diffusion coefficients become overestimated.

However, using the equilibrium crystal divided into octahedral layers is not enough to avoid systematic errors. Firstly, the equilibrium crystal shape differs from the ideal octahedron [19] due to the truncated corners. Secondly, the orientation of the octahedral layers must at all times coincide with the crystal orientation. However, during our prolonged calculations of bulk diffusion the crystals were randomly rotating at high temperatures despite the regular using a procedure of zeroing the angular momentum. These rotations appear to be associated with reactive movement of the cationic sublattice relative to diffusing anions and cations. Unfortunately, we have not been able to offer reliable solutions for the problems of stopping this rotation and changing the orientation of the octahedral layers synchronously to the turns of the crystal. That is why the layers of octahedral shape were not used in this work, and the diffusion coefficient was calculated by the original technique (see the text below), which is not dependent on rotation.

Thirdly, if the layer boundaries are parallel to the rows of particles in the crystal, then the number of particles inside this layer will vary greatly due to the thermal vibrations, resulting in large oscillations of MSD. Moreover, MSD oscillations are affected by the difference in the mobility of particles in the neighbor layers. It is also necessary to take into account fluctuations in the lattice constant.

Finally, the possibility of surface melting (which is observed when using potentials Walker-81 [27]), also limits the applicability of the division into octahedral layers.

Therefore, in order to calculate the diffusion coefficient, we used layers of spherical shape, which are suitable for all the phase states, not bound to the crystal orientation and not parallel to the rows of particles. Investigation of anion diffusion showed [17] that the mobility of the particles in the surface layer is considerably higher than mobility of the rest of the particles, which remains constant through all the inner layers. Therefore we only needed to divide the crystal into two regions: the surface layer (with thickness of one lattice period) and the inner region.

The next important issue it that nanocrystals simulated under IBC have a finite size limiting the maximum value of MSD, while MSD growth is unlimited under PBC. Firstly, this limitation lead to saturation of MSD curves during sufficiently long simulation, when a large number of particles reach the border of the crystal (at least briefly). In order to prevent this saturation, one can regularly update the initial positions of the particles, also redistributing the particles in layers (regions) over again. Secondly, movement of a particle from the surface into the bulk (or the opposite) can lead to an underestimation (overestimation) of DC, calculated for the corresponding region. The largest error occurs when a particle from the bulk comes to the surface, travels a long distance there (due to the increased mobility of the surface particles) and goes back to the bulk. Therefore, it is necessary to exclude from the calculation every particle, which has left its "home" region (i.e. the region with its initial position), still taking into account the distance that it has traversed while remained in that region. However, exclusion of these particles leads to reduction in the number of particles contributing to the MSD, so the amplitude of MSD oscillations grows, and hence the random error increases in time. This effect should also be considered when selecting the rate of updating particle initial positions.

Chaotic rotations mentioned in the discussion of octahedral layers affect the MSD of cations in the nanocrystal greatly, despite the use of spherical boundary between the surface and the bulk, which is not dependent on crystal orientation. Fig. 3 shows that due to such rotations the dependence of MSD on time is not increasing monotonously, but undergoes oscillations of large amplitude. To exclude the effect of rotations, we propose using a threshold filter applied to contribution of individual particles to MSD, i.e. displacement of a particle should be added to MSD only if it has moved far enough – in our case, if the displacement exceeds half of the average distance between neighbor cations. With use of such a filter,

MSD value includes the diffusion hops only, provided that the crystal does not have time to rotate too much (the maximum allowable rotation angle decreases with increasing crystal size). This condition can be satisfied during simulation of any duration, if initial positions of the particles are updated regularly. Fig. 3 shows that without this technique of rotation compensation the MSD could be overestimated by a factor of five (despite the regular zeroing of angular momentum) so that MSD fluctuations would not reflect the diffusion movement of the cations. Due to reactive nature of these chaotic rotations their rate has probably the same exponential dependence on temperature as the diffusion coefficient, so the proposed technique should be used for the cation DC calculation at any temperature.

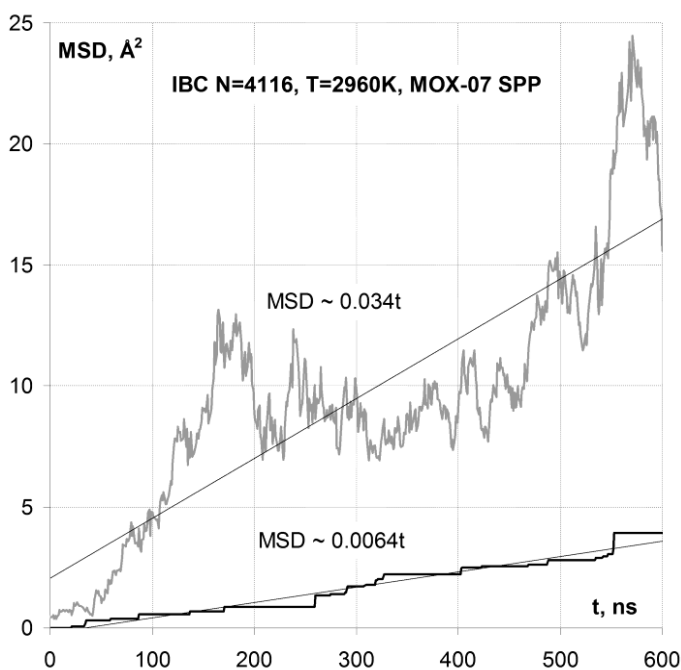

Fig. 3. Time evolution of MSD before (the upper curve) and after (the lower step-wise curve) using of threshold filter, intended to solve the chaotic rotation problem.

## 3. Results and discussion

### 3.1. Calculation of diffusion activation energy using model of point defects

Mass transport in crystalline materials is caused by displacement of atoms away from their equilibrium positions; therefore point defects play an important role in this process [1]. For example, vacancies provide a space which can be occupied by atom from some of the neighbor lattice sites [41] [42] [43], interstitial diffusion mechanism is also widespread.

The diffusion mechanism indicates a way of atom relocation from one position of crystal lattice to another. This relocation is usually interpreted as ion (or several ions) overcoming the energy barrier, so the activation process is considered. The diffusion activation energy of each mechanism can be estimated theoretically via the energies of formation and migration of point defects (for simplicity, at this moment we do not consider more complex defects such as clusters, dislocations, etc.).

The results of calculations of defect formation energies using the method of lattice statics and the density functional theory (DFT) in comparison with the experimental estimates are shown in Table 1. The validity of our original calculation technique is confirmed by comparing the results with the values from review of Govers et al. [37], which were obtained for the four SPPs considered in both works: the differences are 1–5% only.

It is evident that all the calculated energy values of Frenkel defect formation (for both empirical pair potentials and modern DFT-calculations [44] [45]) are more than 50% higher than the experimental estimates. Such an overestimation was mentioned in the earliest works [38] [39] [46]. Note that Morelon-03, MOX-07, Yakub-09 and Basak-03 SPPs have the lowest values for Frenkel defect (15.6, 15.6, 15.9, 16.8 eV, correspondingly).

The calculated energies of Schottky defect formation are also too high, but the lowest values of 7.7–7.8 eV are shown by other SPPs (Walker-81 and Goel-08), which relatively poorly reproduce the thermophysical characteristics of $UO_2$ [26]. The close value of 8.0 eV is shown by more adequate SPP Morelon-03.

Since the calculated energy of the Schottky defect formation is much smaller than that of the Frenkel defect, it has been concluded (see, for example, [47]) that the diffusion of cations in stoichiometric $UO_2$ occurs via Schottky defects, rather than Frenkel defects. Also of interest are the recent static DFT-calculations of Dorado et al. [48], in which the energies of vacancy and interstitial migration of cations are compared (total of five probable scenarios of migration). The authors suggested that interstitial migration mechanism is very unlikely due to its high activation energies over 14 eV, and the most likely mechanism is the "oxygen-assisted" vacancy mechanism, which is significantly influenced by the displacements in the oxygen sublattice.

The activation energy of diffusion in $UO_{2\pm x}$ and other compounds with the fluorite structure is typically estimated using a simple thermodynamic model, proposed by Lidiard and Matzke in 1966 (see the references in [8]). For cation diffusion the corresponding formulas are:

For vacancy diffusion:
$E_A = \Delta G_{SD} - 2 \times \Delta G_{AFD} + \Delta H_{VM}$ in $UO_{2+x}$;
$E_A = \Delta G_{SD} - \Delta G_{AFD} + \Delta H_{VM}$ in $UO_2$; (2)
$E_A = \Delta G_{SD} + \Delta H_{VM}$ in $UO_{2-x}$;
For interstitial diffusion:
$E_A = \Delta G_{FD} - \Delta G_{SD} + \Delta H_{IM}$ in $UO_{2-x}$.

Here, $\Delta G_{SD}$, $\Delta G_{FD}$, $\Delta G_{AFD}$ are the free energies of formation of the corresponding defects, $\Delta H_{VM}$ and $\Delta H_{IM}$ are the enthalpies of the vacancy and interstitial migration. That model predicts a decrease of the activation energy with increasing $x$ for vacancy mechanism and the reverse dependence for interstitial mechanism. The experiments point to the existence of a minimum of DC near $x = -0.02$ at 1600° C, so it is considered that vacancy diffusion mechanism dominates at $x > -0.02$ and interstitial mechanism dominates at $x < -0.02$ [8] [2].

One of the authors of the aforementioned thermodynamic model commented its range of applicability with the following words [8]: "This model is based on simple mass action laws, and it neglects interaction between defects and second phase precipitates ($U_4O_9$ into $UO_{2+x}$, U-metal in $UO_{2-x}$, etc.). It does not allow for the formation of shear structures, for the formation of complexes of oxygen defects in $UO_{2+x}$ and for effects due to valence changes of uranium. Therefore, while it could be expected to be strictly applicable for small deviations from stoichiometry, it will at most show the qualitative behaviour for bigger deviations".

In order to calculate the migration energy, the method of Nudged Elastic Band (NEB) or its modification is usually used [49], where the required energy is determined via the maximum of energy relief along the optimal path.

In this paper, the migration energy was estimated not only by the NEB technique at zero temperature, but also by MD simulation at high temperatures. The crystal simulated under PBC was altered by embedding a defect. The cation diffusion mechanism in this system was determined by the type of the embedded defect, which was confirmed by visual observations. Moreover, in assumption that the concentration of thermal cationic defects is negligibly low compared to the concentration of the embedded defects (i.e. the total concentration does not depend on temperature), the cation diffusion activation energy was supposed to be equal to the vacancy migration energy in the case of the embedded Schottky defect (i.e., trivacancy) and interstitial migration in the case of the embedded interstitial cation (with two interstitial anions maintaining electroneutrality). So, the estimates of these migration energies were obtained by calculating the cation diffusion activation energies from the Arrhenius dependences of diffusion coefficient.

In Table 2 the results of both approaches are compared with the results of DFT-calculations and the experimental estimates.

It is seen that the results of static calculations of the cation migration energy fall within the range of 3.7–5.8 eV for both the vacancy and interstitial migration mechanisms, and these values are 2–6 times higher than the experimental estimates. MD simulation gives values that are closer to the experimental values, and the vacancy migration energy is 2–4 times higher than that of the interstitial. Since MD-estimates take into account the thermal expansion in the temperature range of interest and the disordering of anionic sublattice, in contrast to static calculations at zero temperature, we will analyze high-temperature cation diffusion relying just on them.

In Tables 1–2 we have all the values needed to calculate the diffusion activation energy using formulas (2). The results of this calculation are shown in Table 3. It is seen that even the results calculated using the experimental estimates of the defect formation energies and migration enthalpies differ from the experimental values of diffusion activation energy. The most noticeable difference is in the case of $UO_{2+x}$, where the calculated values of 1.4 eV and 1.9 eV are lower than the Matzke's activation energy of 2.7 eV by 40–80% and 2–3.5 times lower than experimental values of other authors. Note that the difference of Matzke's experiments from other experimental studies can be explained by the fact that his study of $x$ dependence was conducted for mixed oxides $(U_{0.8}, Pu_{0.2})O_{2\pm x}$ [2] and $(U_{0.85}, Pu_{0.15})O_{2\pm x}$ [8]. Plutonium was added in order to obtain a hypostoichiometric crystal, because $U^{4+}$ ions already have the lowest state of oxidation, and plutonium ions are easily reduced to the tri-valent state [8].

Estimates of the diffusion activation energies calculated in the approximation of pair potentials are close to the estimates based on DFT-calculations. However, the difference with the experimental values is satisfactory only in the case of $UO_{2+x}$ (except Basak-03 and Goel-08 SPPs). Overestimation of results for $UO_2$ and $UO_{2-x}$ is mainly due to the high values of the Schottky defect formation energy.

## 3.2. MD simulation of the cation bulk diffusion

### 3.2.1. Quasi-infinite crystals under PBC

Quasi-infinite crystal without artificial defects simulated under PBC is the simplest model for molecular dynamics. It lacks a surface, therefore formation of Schottky defects is prohibited, and only Frenkel disordering occurs. In this paper, we at first calculated the diffusion coefficients of the uranium cations in such a system. It required simulation times of at least 10 ns (two million MD steps), because even for temperatures that are close to the melting point (different for each SPP [27]) the calculated values of DC do not exceed $10^{-7}$ cm$^2$/s (see Fig. 4). For all the five considered SPPs the diffusion activation energy appeared to be greater than 15 eV, so even the increasing of simulation time up to 170 ns allowed to calculate the cation diffusion coefficients in a narrow temperature intervals within 210–450 degrees prior to the melting point only, whereas anion diffusion was earlier studied [4] in wide temperature ranges of 2000–4300 degrees depending on SPP.

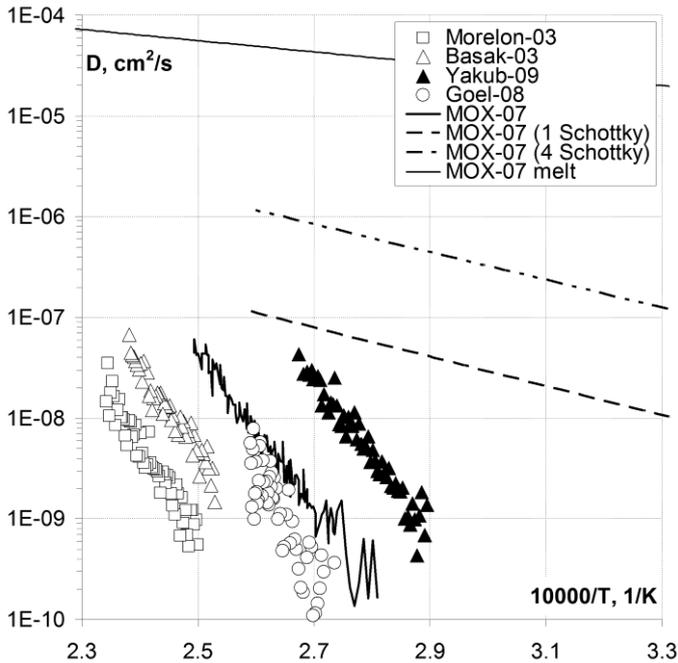

Fig. 4. Temperature dependences of the cation diffusion coefficient obtained by MD simulation under PBC.

The DC dependence for Goel-08 potentials reaches only ~$10^{-8}$ cm$^2$/s before melting, which indicates their lower stability. This is probably connected with the fact that this SPP provided the extremely low melting point [27].

In addition to the results for crystal without artificial defects, we plotted on Fig. 4 the dependences for the melt (with the activation energy of ~1.1 eV) and for systems with one and four Schottky defects (the activation energy of ~2.9 eV).

Based on the temperature dependencies obtained one can say that other authors could not register cation diffusion in the system without artificial defects for two reasons. First, no calculations near the melting temperature of the model crystals have been carried out (see, e.g., [7] [6]) because significant overestimation of the melting point [27] in simulations under PBC compared to experimental values (by at least 600 K) and values obtained by two-phase MD simulations (by at least 500 K) was not taken into account. Second, the temperature step of 100–250 degrees used in the previous works (see the review [4]) was too coarse for studying the diffusion with extremely high activation energy, because each successive simulation would require a dramatic increase of simulation time or system size (as shown in Section 2.2).

All the temperature dependences in Fig. 4 are linear in Arrhenius coordinates, indicating that a single diffusion mechanism persists through this temperature range. Consequently, a single value of the cation diffusion activation energy can be attributed to each SPP in this simple model.

The dominant diffusion mechanism was determined by visual observation of ions movement. As in the case of anions [4], observations showed the absence of long-lived Frenkel pairs regardless of system size. Instead, every time when formation of a Frenkel pair occurred, it caused cyclical permutation of the adjacent cations, concluded by soon recombination of the pair. This mechanism is often called exchange, although some authors distinguish direct exchange of two ions and ring substitution of more than two ions. Exchange diffusion probably occurs in natural crystals, which is supported by experimental observation of short-lived anti-Frenkel pair formation in UO$_2$ and similar crystals CaF$_2$, PbF$_2$ and SrCl$_2$ of fluorite structure [51] [3].

We found that the exchange usually begins with the displacement of two ions, when the first ion acquires additional energy and pushes its neighbor into an interstitial site, as this neighbor starts to move soon after the first ion. If this interstitial site is near the lattice position left by the first ion, then recombination of the Frenkel pair occurs immediately (exchange of two ions, which lasts about 2 ps). Otherwise, ions in turns occupy the vacancy formed, which can also be viewed as the opposite motion of this vacancy. The closeness of interstitial ion and vacancy is energetically favorable, so recombination of the Frenkel pair is almost inevitable, and it occurs soon enough. The instability of the configurations with interstitial ion is accompanied with increased amplitude of movement of the neighbor ions, which sometimes assists concurrent exchanges in other parts of the crystal.

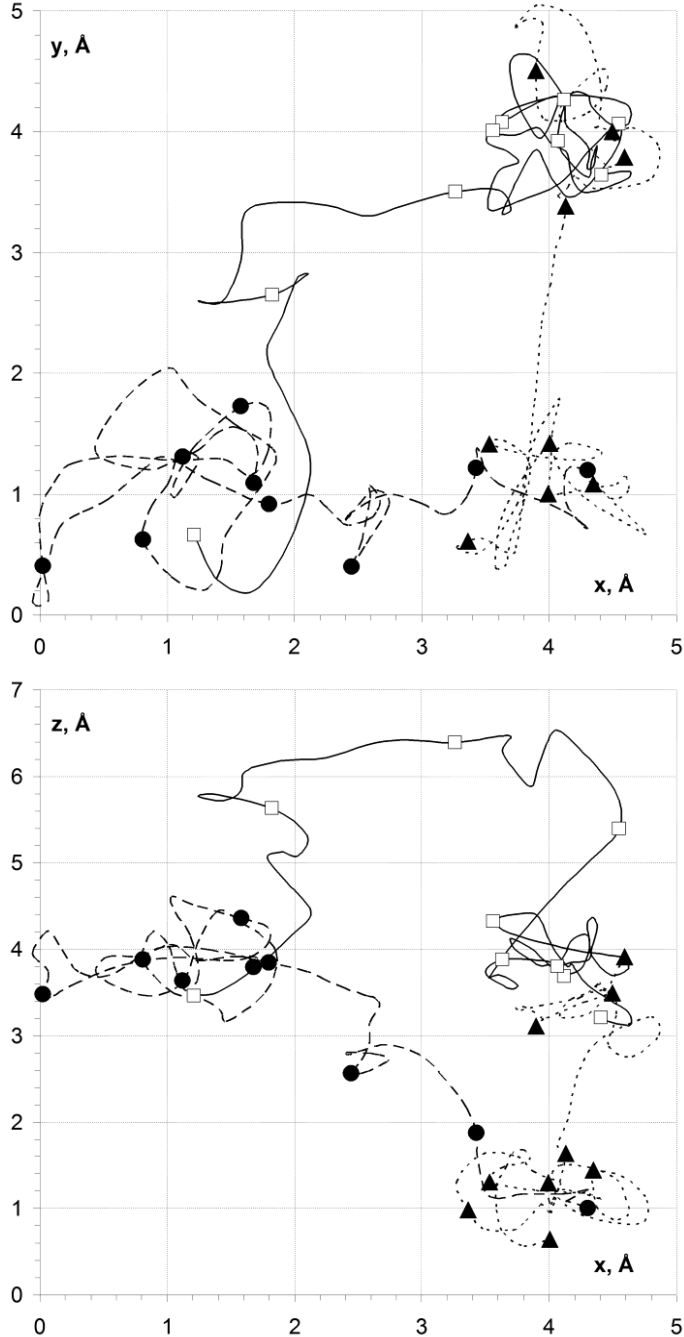

Fig. 5. Projections of trajectories of three cations during sample exchange (one symbol for each hundred MD steps). The movement is clockwise.

An example of such exchange is illustrated by trajectories of three cations in Fig. 5 and their velocities in Fig. 6 (which were calculated from averaged ion positions in order to mitigate thermal oscillations and emphasize motion between lattice sites). The first ion starts to move with sharp acceleration towards an adjacent lattice site (see the steep slope in Fig. 6). The second ion starts to move after 80–100 MD-steps. After another 100 steps the second ion starts to oscillate around the interstitial position (see Fig. 5), while the first ion is displaced further away and already begins to approach the lattice site occupied by the third ion. The third ion, receiving impulse from the first ion, is greatly accelerated and overcomes the distance to the vacancy during 100 steps. At this time, the second ion reaches new lattice site too, while the first ion finished travelling after another 100 MD steps. The whole process lasted 2–2.5 ps.

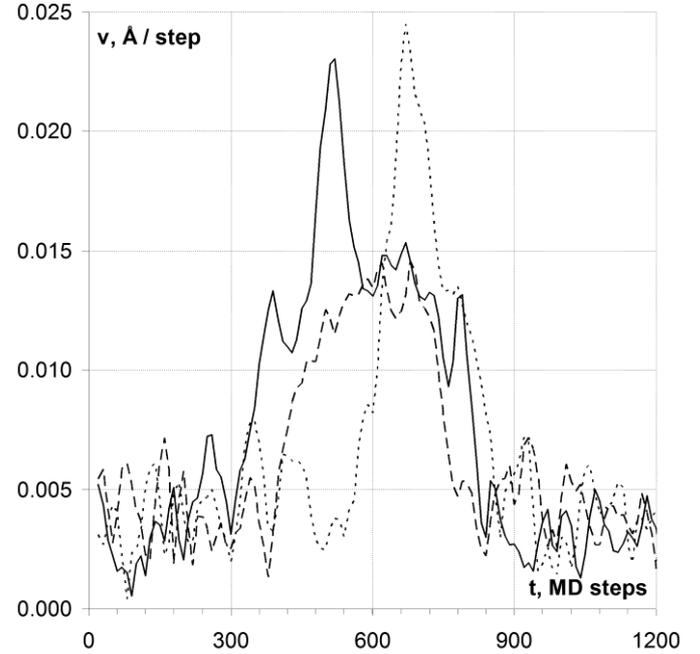

Fig. 6. Velocities of three cations during sample exchange.

Sometimes, additional ions moved to interstitial sites near permutating particles, but afterwards they just returned to their original positions, not participating directly in the cyclical movement. In particular, we encountered an interesting case of five cations exchange, in which five other neighbor cations moved out of their sites to interstices or other sites, and in the end went back. This process lasted 11 ps.

Processing of the temperature dependences of DC gave the diffusion activation energies in the range of 15–22 eV (see Table 4), which for Basak-03, Yakub-09 and MOX–07 SPPs are almost equal to Frenkel defect formation energy, and for Morelon-03 and Goel-08 are approximately 20% above it. This is consistent with our previous results for anions [4], where the activation energy of the exchange diffusion was close to the AFD formation energy. These high activation energies correspond to decrease in DC by an order of magnitude every 150 kelvins, which became a significant obstacle to the calculations near and below the superionic transition temperature (see these temperatures for different SPPs in [26]), where one can expect a change in the slope of the DC dependencies.

As discussed in Section 3.1, the experimental data indicate the dominance of vacancy diffusion of cations in UO$_2$, in which Schottky defects emerges on surface of the crystal and then immerse inside it. However, crystal simulated under PBC has no surface. Therefore, Schottky disordering is prohibited, and only formation of Frenkel pairs occurs. The energy of formation of these pairs for all the considered SPPs exceeds 15 eV; hence, the diffusion activation energy is twice as high as 7.8 eV, the maximum of the experimental values (obtained for (U$_{0.8}$, Pu$_{0.2}$)O$_{1.98}$ [2]). Thereby, periodic boundary conditions seem inappropriate for simulation of uranium self-diffusion (as in the case of oxygen [4]) in stoichiometric UO$_2$ crystals at low temperatures up to the superionic transition.

However, the study of self-diffusion in the anionic sublattice showed [4], that the exchange diffusion mechanism occurring under PBC becomes dominant at high temperatures near the superionic transition, regardless of the presence of surface (and this might be true for real crystals as well). So, it can be supposed that this mechanism of transport also becomes dominant in the cationic sublattice prior to the melting. Thus, the direct measurement of cation self-diffusion in the similar compound CaF$_2$ indicates a sharp increase in their DC after superionic transition [52], which can be an evidence of transition to the exchange diffusion mechanism. Therefore, the use of PBC for simulation at high temperatures near the melting point could have a practical sense.

It is significant that direct comparison of the calculations discussed above with the experimental data is not possible at this moment. The existing experimental data were measured at the temperatures below 2300 K for imperfect crystals (containing impurities, dislocations and grain boundaries) with surface, while our model diffusion coefficients were obtained under PBC (i.e. without surface) for temperatures above the experimental melting point of UO$_2$. Direct comparison requires obtaining experimental data at higher temperatures (up to the melting point) or simulating crystals with a free surface at lower temperatures, because such crystals have the melting temperature close to the experiment [27] and feature the diffusion of cations accelerated by the vacancies immersed from the surface.

Fig. 4 also shows that the cation diffusion coefficient increases by almost three orders of magnitude at the melting point. In the melt, a free movement of cations was observed with activation energy of 1.03–1.15 eV (see Table 5). Moreover, the DC temperature dependences for the bulk and surface cations in the melt are virtually the same and are not sensitive to the system size. Differences related to the choice of SPP in the melt appear to be relatively small.

### 3.2.2. Nanocrystals with free surface

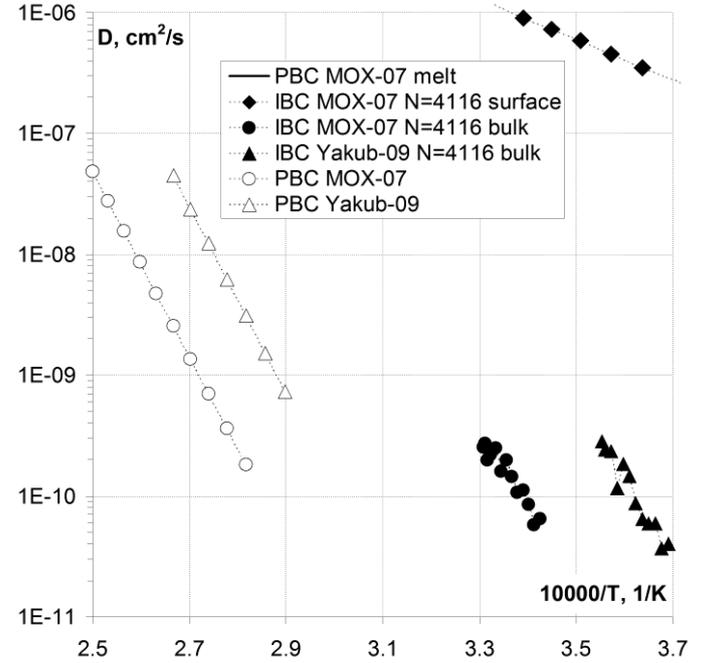

Fig. 7. Cation self-diffusion coefficients, obtained by MD simulation under PBC and IBC (both in bulk and on surface of nanocrystals). The experimental data of Matzke are extrapolated to the melting point.

For the consumptive MD calculations of cation self-diffusion in the bulk of nanocrystals surrounded by vacuum we chose two SPPs, Yakub-09 and MOX-07, which reproduced the other characteristics of UO$_2$ better than the rest (see [26] [27] [4]). As a result we obtained the following temperature dependences for UO$_{2.00}$ nanocrystals (see Fig. 7):

$$D_{MOX-07} = (2.86^{+127.2}_{-2.80}) \times 10^8 \exp\left(-\frac{10.8 \pm 1.0 \, eV}{kT}\right),$$
cm$^2$/sec, $2920K \leq T \leq 3023K$

$$D_{Yakub-09} = (3.62^{+407.1}_{-3.58}) \times 10^{13} \exp\left(-\frac{12.9 \pm 1.1 eV}{kT}\right),$$
cm$^2$/sec, $2710K \leq T \leq 2815K$ (3)

The higher uncertainty of coefficients of these dependencies, compared with the results for PBC, is due to smaller temperature range of only 100 K. The lower temperatures are currently inaccessible, because it requires simulation time of over 2000 ns (400 million MD-steps) to calculate the DC of cations for them. The highest DCs obtained under IBC (for nanocrystals of 4116 ions) did not exceed $10^{-9}$ cm$^2$/s and required simulation time of about 200 ns to be measured. The activation energies of both dependences (3) are considerably lower than the activation energies of exchange diffusion in the quasi-infinite crystals discussed above (see Table 4), but still above the theoretical estimates of the vacancy diffusion activation energy of 8.7 eV and 8.4 eV for these potentials (Table 3). Hence, a competition between two (or more) diffusion mechanisms can be assumed in this temperature range. Visual observation showed that cations diffuse via either the exchange (with Frenkel disordering, see Fig. 5) or the vacancy (with Schottky disordering, see Section 3.2.4) diffusion mechanism. Nevertheless, even at the highest temperatures the first mechanism (exchange) occurred much less frequently than the second (with formation of the cation vacancies on the free surface).

If the diffusion activation energy near the melting point actually reached ~11 eV in real $UO_2$ crystals, it would mean that linear extrapolations of the available experimental dependences measured at T < 2300 K (see Fig. 8) to the melting point underestimate DC by several orders of magnitude (predicting $10^{-13}$–$10^{-10}$ cm$^2$/s [53] [2] [11] instead of $10^{-8}$–$10^{-7}$ cm$^2$/s, obtained in this work). This result should be taken into account when predicting the nuclear fuel behavior near the melting point.

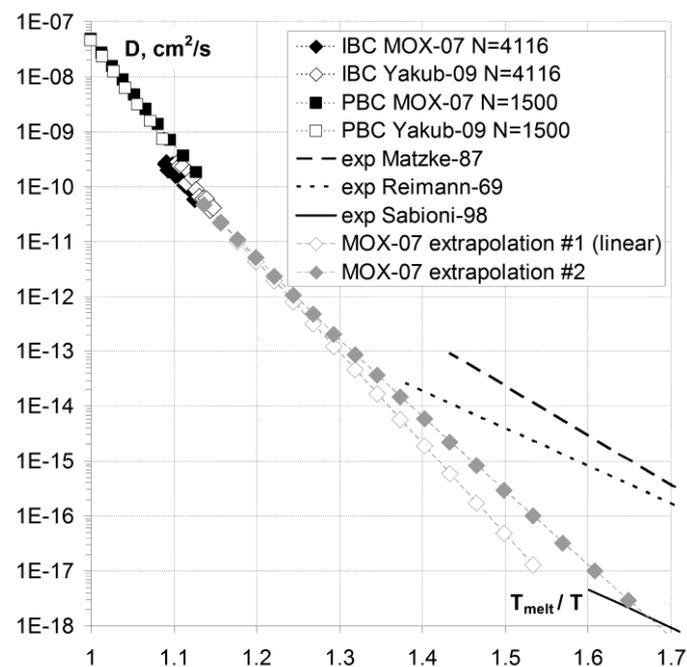

Fig. 8. Dependences of cation self-diffusion on temperature reduced to the melting points. The second extrapolation is based on the phenomenological analysis from Section 3.2.4.

Before a strict comparison of our results with experimental data, let us consider one feature of the calculated DC temperature dependences shown in Fig. 7. It can be seen that regardless of the boundary conditions and SPP these curves are almost parallel to each other and differ substantially by their starting points only. In particular, the highest value of DC under PBC is about $5\times10^{-8}$ cm$^2$/s for several SPPs, while the highest value of DC under IBC is two orders of magnitude smaller ($3\times10^{-10}$ cm$^2$/s). It appears that the shift of the dependences on the temperature scale is related to the difference in the melting points for the corresponding model crystals, and the shift along the DC axis with the change of the boundary conditions can be matched to the dependence of the melting temperature on the nanocrystal size studied in [27].

Similar observations have been made previously during analysis of the experimental data. Thus, Matzke [54] noted that the DC temperature dependences for cations in several compounds with the fluorite structure ($UO_2$, $ThO_2$, $CaF_2$) become close to each other when using the scale of $T_{melt}/T$ instead of $10^4/T$. Later [2] Matzke showed that such correlation between DC and $T_{melt}$ holds true for both sublattices (metallic and non-metallic) by adding to single chart the dependences for anions in $UO_2$, $ThO_2$, $PuO_2$, $CaF_2$ and $BaF_2$, and also dependences for anions and cations in $Li_2O$ with the anti-fluorite structure.

Our model nanocrystals of $UO_2$ also melt at different temperatures depending on the SPP, and the use of PBC increases melting temperature by more than 20% [27]. Fig. 8 shows that the use of the temperature scale reduced to $T_{melt}$ (for IBC-dependences the results of the parabolic extrapolation of $T_{melt}$ to macrocrystals from [27] were used) brings together not only curves for different potentials, but also the results for PBC and IBC.

Since this empirical relationship holds true regardless of the atom type in compound and simulation conditions, it is only correct to compare the results of our simulations with the experimental data after this reduction. At the comparison of "DC vs. $T_{melt}/T$" curves for several compounds in the review [55] it was also noted that the extrapolations of dependences tend to converge to values of the order of $10^{-8}$–$10^{-9}$ cm$^2$/s at $T_{melt}$ (when considering the slower species, i.e. U in $UO_2$ and O in $Li_2O$). This limit matches to our simulations under IBC, whereas DC calculated using PBC achieve higher values ~$10^{-7}$ cm$^2$/s due to the overheating in the absence of the surface. Now, let us remind that linear extrapolations of the experimental data on cation diffusion in $UO_2$ [53] [2] [11] reach $10^{-13}$–$10^{-10}$ cm$^2$/s at the melting point, i.e. the values of $10^{-8}$–$10^{-9}$ cm$^2$/s are reachable only if there is a change in the diffusion activation energy. Moreover, such a change was detected experimentally for cations in $CaF_2$ [52]. Thus, our temperature dependences do not contradict the experimental data, but indicate the possibility of change in the diffusion mechanism.

One of the reasons for decrease in the effective diffusion activation energy with decreasing temperature is the presence of temperature-independent defects such as impurities, dislocations or grain boundaries, because these defects can significantly reduce ore exclude contribution of defect formation to the effective diffusion activation energy. In the most recent experimental work on cation diffusion measurement in single crystals of $UO_2$ [11] the absence of grain boundaries and concentration of impurities were carefully checked, but the presence of dislocations was not discussed. In addition to lowering the activation energy, this hypothesis can also explain the fact that the uranium diffusion coefficients in different studies vary by several orders of magnitude: their samples may differ by the temperature-independent defect concentration.

The effect of impurities on cation diffusion was measured experimentally in uranium carbide [56]. A small concentration of atoms of other metals (120 ppm) caused a bend on the DC curve where the diffusion activation energy decreased from 6.13 eV down to 3.65 eV at a temperature of about 2370 K (~90% of the melting point), while the purer sample (with impurities of less than 30 ppm) retained activation energy of 6.17 eV over the temperature range of 1800–2500 K. In this context it is interesting that the lowest cation DCs were obtained for $UO_2$ single crystals with larger impurity concentration (not less than 200 parts per million) [11].

Another reason for lowering the activation energy in the experiment could be presence of reduced cations $U^{3+}$, which were absent in our model. These cations can have migration energy decreased by ~2 eV, as shown by the static calculations of Jackson and others [47].

### 3.2.3. Cation diffusion in non-stoichiometric uranium dioxide

In order to evaluate the effect of temperature-independent defects of given concentration on the diffusion activation energy we have simulated the non-stoichiometric uranium dioxide crystals both oxygen-excess and oxygen-deficient.

Fission of uranium nuclei and oxygen redistribution lead to increase in the stoichiometry in the center of a fuel rod up to approximately $UO_{2.08}$, and in order to reduce the fuel-cladding chemical interaction of MOX-fuel its stoichiometry, on the contrary, is reduced beforehand, and the following ions redistribution lower the O/M ratio to 1.92 in the rod center [57].

Change of the O/M ratio should be compensated by change in the charges of ions in order to maintain electroneutrality of the crystal. In our calculations under PBC violation of electroneutrality only increases the mobility of ions and decreases the melting temperature, but nanocrystals simulated under IBC eject "excess" ions to the surrounding space.

In the literature two ways of maintaining electroneutrality of model $UO_{2\pm x}$ crystals have been proposed. The first method [58] [59] [57] is replacement of $U^{4+}$ ions with $U^{5+}$ ions near interstitial anions (in $UO_{2+x}$) and with $Pu^{3+}$ ions near anion vacancies (in $(U_{0.8}, Pu_{0.2})O_{2-x}$). Additional parameters of interatomic potentials were fitted there to the lattice constant of $UO_{2+x}$ and $Pu_2O_3$, respectively. In the second method [32] interatomic potential $U^{5+}$–$O^{2-}$ fitted to the lattice constant of the β-phase $U_4O_9$, and mobility of polarons at high temperatures were taken into account via the special algorithm of regular hopping of electron hole (i.e., hopping of oxidized ion $U^{5+}$), however, the hopping characteristics were chosen constant over all the temperatures studied.

In this paper an even simpler approximation is chosen, in which the electroneutrality is compensated by uniform change in charge of all anions. Similar to the partially ionic potential model generally used for stoichiometric $UO_2$ simulations, this approximation corresponds to the choice of some average degree of electronic subsystem disordering. We assumed that it would be good enough to describe the state of the system near the melting point due to delocalization of polarons. Since we did not change the short-range part of the SPP, the adequacy of this model should decrease with increasing deviation from stoichiometry. However, we hope that this approach can provide qualitatively correct conclusions.

It is known [60] that cationic sublattice of uranium dioxide almost does not change with deviation from stoichiometry; instead there are significant changes in the layout of oxygen ions. It order to verify the validity of our approximation using this fact, we formed crystals of $UO_{2.10}$ and $UO_{2.15}$ on the basis of $UO_{2.00}$ crystal (of 4116 ions) by removing 64 and 96 random cations, respectively. As expected, these cation vacancies shortly came to the surface in the course of MD simulation at the temperature of 2900 K, while the equilibrium number of interstitial anions became distributed through volume of the crystal.

Another test is determination of the melting temperature of these relaxed nanocrystals. The hyperstoichiometric nanocrystals of $UO_{2.10}$ and $UO_{2.15}$ melted at a temperature 30 and 50 degrees lower than $UO_{2.00}$ nanocrystals, which roughly corresponds to the experimental data [61]. Unfortunately, the melting point of the hypostoichiometric crystal $UO_{1.85}$ was 30 degrees higher, while experiments indicated the decrease. We assume that this deviation does not affect the cation diffusion mechanism, although in order to provide a quantitative accuracy the correction of pair potentials taking into account the stoichiometry deviation is needed.

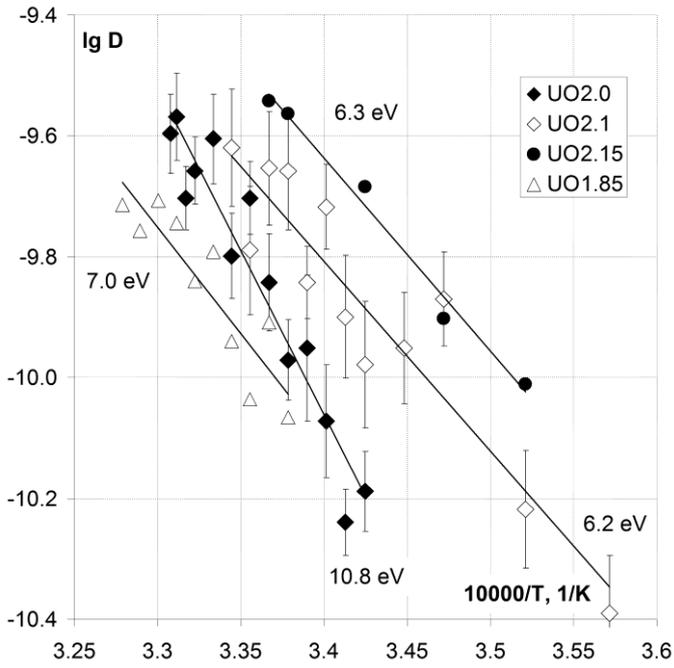

Fig. 9. Cation self-diffusion coefficients in bulk of non-stoichiometric uranium dioxide nanocrystals.

For cation diffusion in $UO_{2\pm x}$, we obtained the following temperature dependences (see Fig. 9):

$$D_{MOX-07}(UO_{1.85}) = 71^{+7425}_{-70}\exp\left(-\frac{7.0\pm1.2\,eV}{kT}\right);$$
$cm^2/sec$, $2960K \leq T \leq 3050K$

$$D_{MOX-07}(UO_{2.10}) = 7.5^{+155}_{-7.1}\exp\left(-\frac{6.2\pm0.8\,eV}{kT}\right);\quad (4)$$
$cm^2/sec$, $2800K \leq T \leq 2990K$

$$D_{MOX-07}(UO_{2.15}) = 17^{+82}_{-14}\exp\left(-\frac{6.3\pm1.3\,eV}{kT}\right).$$
$cm^2/sec$, $2840K \leq T \leq 2970K$

It can be seen that when $x > 0$ increasing deviation from stoichiometry leads to an increase in DC at the same temperature. The activation energy of diffusion in $UO_{2.10}$ and $UO_{2.15}$ is almost equal (as in the experiments [62] [8]), but it is 4.5–4.6 eV lower compared to $UO_{2.00}$. The formulas (2) also predict lowering of the activation energy by the value of anti-Frenkel defect formation energy (4.1 eV in the case of MOX-07 potentials), but in the experiments the difference of activation energies in $UO_{2.00}$ and $UO_{2+x}$ is considerably lower (from 0 to 3 eV according to different authors, see Table 3).

Interestingly, the activation energy of diffusion in $UO_{1.85}$ is near those for $UO_{2.10}$ and $UO_{2.15}$ (in contrast to the experimental data), while the shift of the DC dependence to higher temperatures caused its crossing with the curve for $UO_{2.00}$ (see Fig. 9).

In general, the strong influence of temperature-independent anionic defects on the diffusion activation energy, confirmed by our simulation of non-stoichiometric uranium dioxide, allow explaining the large scatter of the available experimental dependences [53] [55] [2] [11] by lack of continuous control of stoichiometry and different concentration of temperature-independent of all types.

Visual observations in all the cases of stoichiometry deviation showed the dominance of the vacancy diffusion mechanism, which contradicts the general hypothesis stating the transition to interstitial mechanism at $x < -0.02$ [2] (the mass transport of Schottky trios could not be excluded though [8]). Nevertheless, discussing that hypothesis Catlow noted [38]: "However, it is of course possible that some mechanism other than interstitial migration becomes operative in hypostoichiometric $UO_2$. For instance, the vacancy clustering which is known to occur in other anion-deficient fluorite oxides (see, for example, [63]), could result in low-energy mechanisms for cation migration". So in the next section we will try to identify the similarities and differences of vacancy diffusion occurring in the model nanocrystals of $UO_{2.00}$, $UO_{2+x}$ and $UO_{2-x}$.

### 3.2.4. Clarification of the vacancy diffusion mechanism in $UO_{2\pm x}$

In studying the vacancy migration by static calculations there has been a doubt [64] if the cation vacancy moves to the nearest cation site or chooses diagonal path through an interstitial site. Those calculations showed that the shortest path is more favorable, but the authors considered that fact as counterintuitive. We decided to determine which of the two paths was preferred in our MD simulations. Analyzing trajectories of cations in several cases of the exchange and vacancy diffusion we found that the shortest path (having length of ~0.7 of the lattice constant) was preferred every time.

Observations of the cationic sublattice showed the same mechanism of vacancy movement in $UO_{2.00}$, $UO_{2+x}$ and $UO_{2-x}$: at some time a vacancy formed on the surface immerses into the crystal bulk, travels there for some time, and then returns to the surface, where it ceases to exist (see Fig. 10). Therefore, the differences in the cation diffusion characteristics of $UO_{2.00}$, $UO_{2+x}$ and $UO_{2-x}$ should originate from the structure of the anionic sublattice.

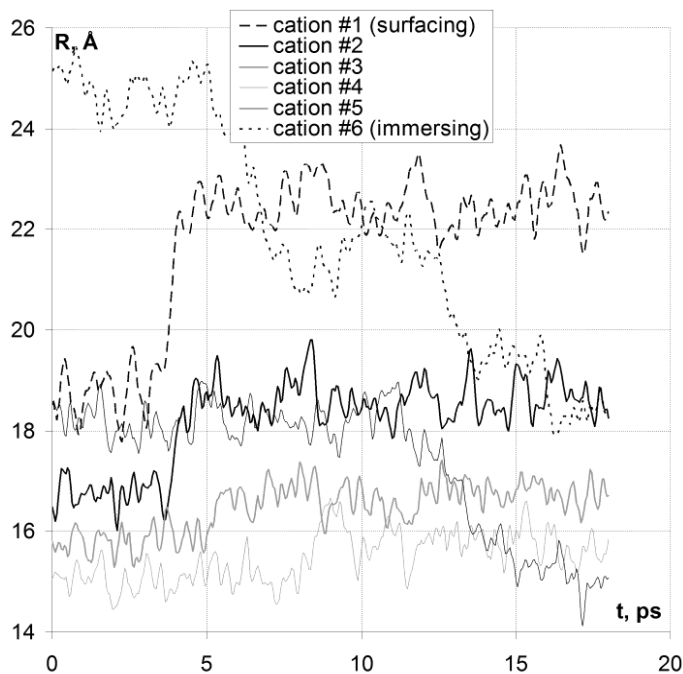

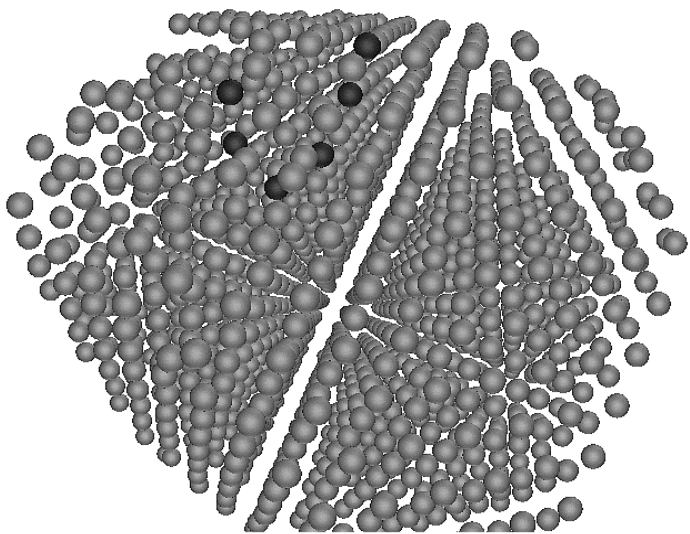

Fig. 10. a) Time evolution of the distance to the crystal center of cations participated in vacancy migration. Migration of cation #1 resulted in formation of the vacancy, while migration of cation #6 resulted in its recombination; b) the path of vacancy migration in cationic sublattice.

The condition of electroneutrality of the crystal bulk implies the cation vacancy in $UO_2$ to be formed together with the ionic defects of the opposite effective charge, such as interstitial cations (the Frenkel disordering) or two anion vacancies (Schottky disordering). Our simulations showed that in the model crystals the second type of disordering causes the vacancy diffusion.

In the theoretical static calculations (as in Section 3.1) the formation energy of the Schottky defect is generally attributed to the neutral trio of one cation and two anion vacancies, which do not interact with each other (presumable, being separated by infinite distance) [38] [37]. In order to assess applicability of this scheme to our model, we first tried to estimate the number of anion vacancies and their distance from the cation vacancy using MD simulations of $UO_2$, $UO_{2.15}$ and $UO_{1.85}$ nanocrystals near their melting points. At such high temperatures (~3000 K) the anionic sublattice is almost completely disordered, which greatly complicates tracking of individual trajectories of all anions traveling near a cation vacancy. Instead, we determined the coordinates of lattice sites, which had been occupied by cationic vacancy, and monitored time evolution of the number of anions within a specified distance from these sites. When cation vacancy entered the next lattice site it induced a minimum on the corresponding time dependence. Thus, subtracting a value of this minimum from a value in the absence of the cation vacancy, we were able to estimate the number of accompanying anion vacancies.

The quantity introduced above appeared to be continuous function of the distance at which the number of the nearest anions was counted, but in all the cases ($UO_{2.00}$, $UO_{2.15}$ and $UO_{1.85}$) its peak was at a distance of about 0.5 of the lattice period. This distance is somewhat greater than the average distance between cation and anion in the ideal lattice equal to ~0.433 of the lattice period.

Fig. 11 shows two examples of such time dependences; each curve corresponds to a single site of the cationic sublattice. It can be seen that the estimated number of anionic vacancies accompanying the cation vacancy in $UO_{2.00}$ is (as expected) close to two, but in $UO_{2.15}$ it is greater by two than in $UO_{2.00}$, and the value for $UO_{1.85}$ almost equals the value for $UO_{2.15}$. This counterintuitive overestimation raises a question about positions of four anion vacancies relative to the cation vacancy and still do not reveal any difference between $UO_{1.85}$ and $UO_{2.15}$.

To clarify the situation further, we tried to determine the configuration of the anionic sublattice near cation vacancy at sufficiently low temperatures, where anions are not so mobile. After immersing of cation vacancy into the crystal we lowered the temperature to 2000 K and equilibrated the system over a few nanoseconds. After that, the temperature was lowered further to 1000 K. This two-step cooling ensured that anions had time to take equilibrium positions, as at 1000 K they almost do not move.

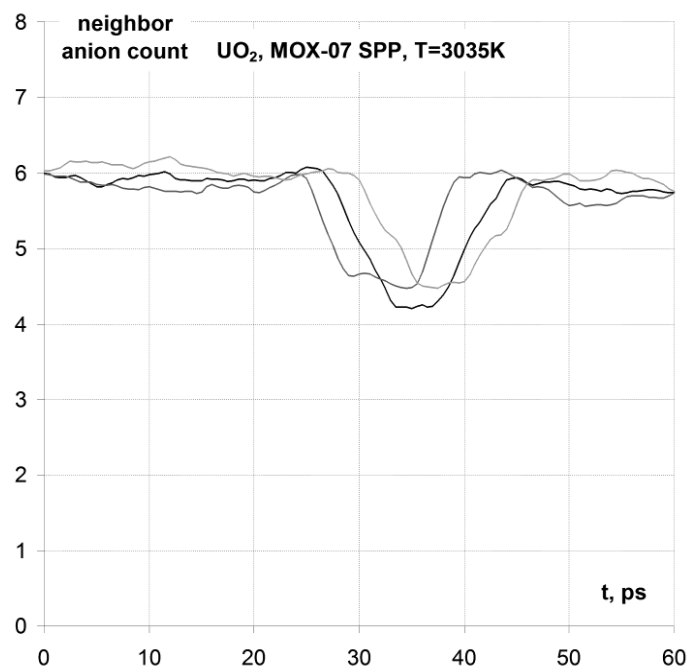

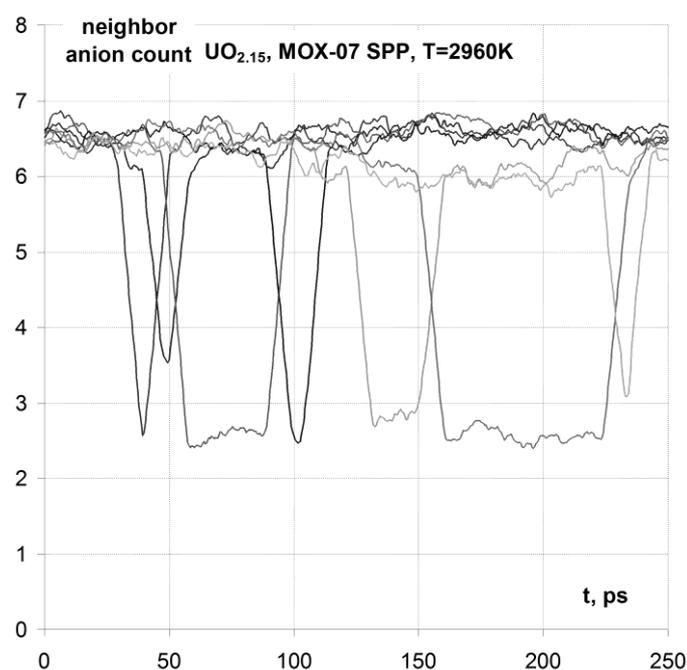

Fig. 11. Time evolution of number of anions near the sites of cationic sublattice, which are visited by cation vacancy.

After examining the structure of anionic sublattice, we found that in $UO_{2.00}$ only one anion vacancy is present near the cation vacancy (see Fig. 12), while the electrostatic potential of the cation vacancy is additionally compensated by deflection of the neighbor ions. The other simulated configurations where the cation vacancy was accompanied by two or zero anion vacancies had excess energy and after some time they also converged to the configuration with one accompanying anion vacancy. This configuration is confirmed by DFT calculations [22], where the divacancy had lower migration energy than the trivacancy migration energy; however their formation energies were almost equal.

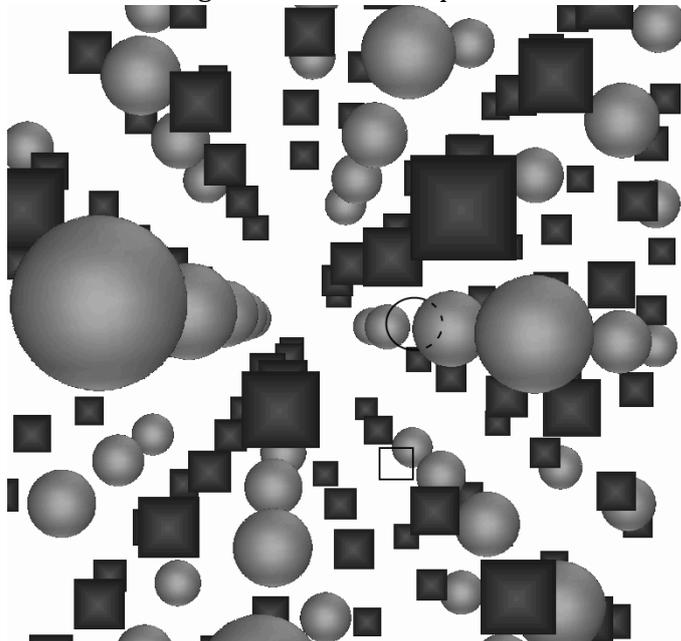

Fig. 12. Cation vacancy (empty circle) accompanied by single anion vacancy (empty square) in a spherical region of the $UO_{2.00}$ nanocrystal.

In $UO_{2.15}$ we observed the absence of classical anion vacancies near the cation vacancy; instead the eight nearest anions formed a perfect cube with an edge length of ~0.79 of the lattice constant (80% longer than usual O–O distance in the defectless $UO_{2.00}$ crystal and 30% longer than edge of the similar cube around single cation vacancy in $UO_{2.00}$). Artificially embedded anion vacancies immediately recombined with the nearest interstitial anions (the picture is omitted). This also correlates with the work of Andersson et al. [22], where a single uranium vacancy in $UO_{2+x}$ was energetically favorable compared to divacancy and trivacancy.

In $UO_{1.85}$ there were four classical anion vacancies near the cation vacancy, which confirmed the above estimate of their number for high temperatures. And the excess positive charge of the cluster of anion vacancies was also compensated by attraction of neighbor anions to the closer distance and repulsion of neighbor cations. In [22] the trivacancy in $UO_{2-x}$ was energetically favorable, but the authors had not considered the possibility of more than two anion vacancies, so that work does not contradict with out results.

Using the new information about configuration of the point defects, we deduced within the model of Lidiard-Matzke the following formulas for the

effective activation energies of the vacancy migration of cations:

$E_A = \Delta G_{SD} - \Delta G_{AFD} \times 2 + \Delta H_{VM}$ in $UO_{2.15}$;
$E_A = \Delta G_{SD} - \Delta G_{AFD}/2 - \Delta H_{B1} + \Delta H_{VM}$ in $UO_{2.00}$;   (5)
$E_A = \Delta G_{SD} - \Delta H_{B4} + \Delta H_{VM}$ in $UO_{1.85}$;

Equations (5) correspond to the crystalline uranium dioxide without the superionic transition.

The equation for $UO_{2.15}$ coincides with the formula of Matzke [Matzke73] [Matzke87], as he did not assumed clustering of vacancies, and according to our results cation vacancies in $UO_{2.15}$ are not bound to anion vacancies (the lifetime of anion vacancies is short due to excess interstitial anions). In the model nanocrystals of $UO_{2.00}$ the divacancy was energetically favorable, so the new formula takes into account the binding energy of cation vacancy with anion vacancy $\Delta H_{B1}$ and the energy of anion vacancy formation ($\Delta G_{AFD}/2$). In the formula for $UO_{1.85}$ the binding energy of cation vacancy with four anion vacancies $\Delta H_{B4}$ is subtracted, while the energy of anionic vacancy formation is excluded from the equation, as the concentration of anionic vacancies in $UO_{1.85}$ is independent of temperature.

Using formulas (5) for estimation of the uranium diffusion activation energy, we assume that the vacancy migration enthalpy $\Delta H_{VM}$ is not significantly dependent on the configuration or number of adjacent anion vacancies, since the calculations for $UO_{2.00}$ confirmed this. The enthalpy $\Delta H_{VM}$ may also depend on the stoichiometry, but for simplicity we have neglected this dependence.

Since formulas (5) describe the crystals without the superionic transition, the values obtained from (5) are appropriate for comparison with the experimental data obtained at relatively low temperatures (below 2300 K).

With the set of pair potentials MOX-07 we obtained the values of 2.06 eV and 5.59 eV for the binding energies $\Delta H_{B1}$ and $\Delta H_{B4}$ using the lattice statics method, while for migration enthalpy $\Delta H_{VM}$ the MD-estimate of 2.9 eV was taken. Correspondingly, formulas (5) give for MOX-07 SPP the following results: $E_A[UO_{2.15}]$ = 9.8 – 4.1×2 + 2.9 = 4.5 eV, $E_A[UO_{2.00}]$ = 9.8 – 4.1/2 – 2.06 + 2.9 ≈ 8.6 eV, $E_A[UO_{1.85}]$ = 9.8 – 5.59 + 2.9 ≈ 7.1 eV. All three estimates are 2–3 eV higher than the experimental values of Matzke (2.6 eV, 5.6 eV, 5.0 eV, respectively), which can be related to the overestimated energy of the Schottky defect formation (see Section 3.1) or influence of various temperature-independent defects in the experiments. However on a qualitative level the model correctly predicts a decrease in the activation energy with increasing deviation from stoichiometry.

For Yakub-09 potentials the formulas (5) give: $E_A[UO_{2.15}]$ = 10.9 – 5.6 × 2 + 3.1 = 2.8 eV, $E_A[UO_{2.00}]$ = 10.9 – 5.6 / 2 – 2.73 + 3.1 ≈ 8.5 eV, $E_A[UO_{1.85}]$ = 10.9 – 6.93 + 3.1 ≈ 7.1 eV. The values of $E_A[UO_{2.00}]$ and $E_A[UO_{1.85}]$ are close to the values for MOX-07 SPP, but the value of $E_A[UO_{2.15}]$ is almost equal to the experimental value Matzke (2.6 eV). In the latter case, the contributions of $\Delta G_{SD}$ = 10.9 eV and $\Delta G_{AFD} \times 2$ = 11.2 eV to $E_A[UO_{2.15}]$, having the opposite sign in (5), almost completely neutralize each other, which probably occurs in the real crystals. Moreover, it leads to an interesting conclusion, that in the crystalline phase of $UO_{2.15}$ the temperature dependence of the cation vacancy concentration could be ceased.

In order to describe the results of high-temperature MD simulation (3)–(4), the relations (5) are to be modified with consideration of the superionic transition. The temperature dependence of the anionic defects concentration, which is exponential in the crystalline phase, weakens or even disappears during this phase transition due to decreasing of their effective formation energy. In the approximation of constant concentration of the anionic defects, the diffusion activation energy of both anions and cations does not depend on the energy of the anionic sublattice disordering. This reasoning is confirmed by our previous work [4], where the effective activation energy of anion diffusion gradually decreased with increasing temperature and reached saturation in the superionic phase, where the concentration of anionic defects reached the maximum value. Therefore, for $UO_{2.00}$ and $UO_{2+x}$ we also expect a smooth change in the cation diffusion activation energy with the temperature in the region of superionic transition. In $UO_{1.85}$ this dependence is almost absent due to the negligible concentration of anti-Frenkel defects.

Exclusion of the anti-Frenkel defect formation energy from (5) allowed calculating the estimates of the effective activation energy of cation diffusion in the superionic phase. For MOX-07 SPP this estimate is $E_A[UO_{2.00}]$ = 9.8 – 2.06 + 2.9 ≈ 10.6 eV, which coincides with the activation energy of 10.8 eV of the temperature dependence (3). In hyperstoichiometric model crystals $UO_{2.10}$ and $UO_{2.15}$ the maximum concentration of anti-Frenkel defects is lower than in $UO_{2.00}$ because anions already occupied a fraction of available interstitial positions. Hence, their effective formation energy is not completely ceased at the melting temperature. The difference between the value of 4.5 eV obtained from the formulas (5), and the values of 6.2 eV and 6.3 eV obtained for $UO_{2.10}$ and $UO_{2.15}$ by MD simulation suggests reduction of the effective energy of anion vacancy formation by 6.3 – 4.5 = 1.8 eV. Finally, the estimate of 7.1 eV obtained from (5) for $UO_{1.85}$ coincided with the activation energy of 7.0 eV of the temperature dependence (4). In the latter case, the superionic transition does not affect the formula.

In the case of Yakub-09 potentials, the small difference between the estimate of $E_A[UO_{2.00}]$ = 10.9 – 2.73 + 3.1 ≈ 11.3 eV for the superionic phase and the activation energy of 12.9 eV of the temperature dependence (3) can be explained by contribution of the exchange diffusion of cations.

Thus, the values of diffusion activation energy from Sections 3.2.2 and 3.2.3 correspond to the phenomenological analysis (5) within the Lidiard-Matzke model. This analysis showed that in the hyperstoichiometric and stoichiometric crystals the high-temperature diffusion activation energy should be greater than at the lower temperatures due to the superionic transition. This effect can manifest itself in the real crystals of uranium dioxide. The increase in the activation energies at the high temperatures is predicted to be about 2 eV. The corresponding extrapolation of our DC temperature dependence for MOX-07 SPP to the region of the lower temperatures is shown in Fig. 8. It can be seen that the extrapolations of the experimental data and the model converge at 60–70% of the melting temperature near the superionic transition.

### 3.3. MD simulation of the surface diffusion of cations

Mass transport at the solid surface (or the surface diffusion) takes part in many processes, such as adsorption and desorption (in particular, the segregation of fission products), heterogeneous catalysis, and crystal growth, wetting etc. [65]. The recent study [25] showed that the movement of bubbles/cavities in $UO_2$ in the presence of a temperature gradient is also determined by surface diffusion.

Matzke had processed the results of nine experimental studies and recommended for the surface diffusion the following temperature dependence [56]:

$$D_S(T) = 5 \times 10^5 \exp\left(-\frac{4.7\, eV}{kT}\right),\quad (6)$$

$1200°C \leq T \leq 1800°C$

Ions on a crystal surface diffuse faster than ions in the bulk as they are less bound (due to less number of neighbors). However, surface particles can have different number of neighbors, which depends on their position: particles at the crystal edges have fewer neighbors than particles at faces, and particles at vertices have even fewer. Our visual observations confirm higher mobility of particles at vertices and edges of a nanocrystal. Moreover, ion at a vertex often remains at the same vertex after movement, as does an ion at an edge. Therefore, trajectories of such particles become closed shortly. After that their MSD ceases to increase and just oscillates near some mean value.

As a result, MSD dependence calculated for all surface particles (in this case all surface cations) is not a straight but a bent line. This effect is most noticeable on smaller crystals (since their fraction of particles at edges and vertices is still quite large): for example, see Fig. 13 with MSD dependence for a system of 768 ions. Its slope through the first three nanoseconds is an order of magnitude higher than the slope, established after 7 ns simulation, because of the ceased effect of particles at vertices and edges. Time of reaching a constant slope depends on the system size, which determines the length of edges, so it should be taken into account when calculating DC and comparing them between nanocrystals of different sizes. Fig. 1 shows dependences obtained without taking into account this effect, so the curves for the smallest crystals of 768 and 1500 ions are noticeably higher than the rest.

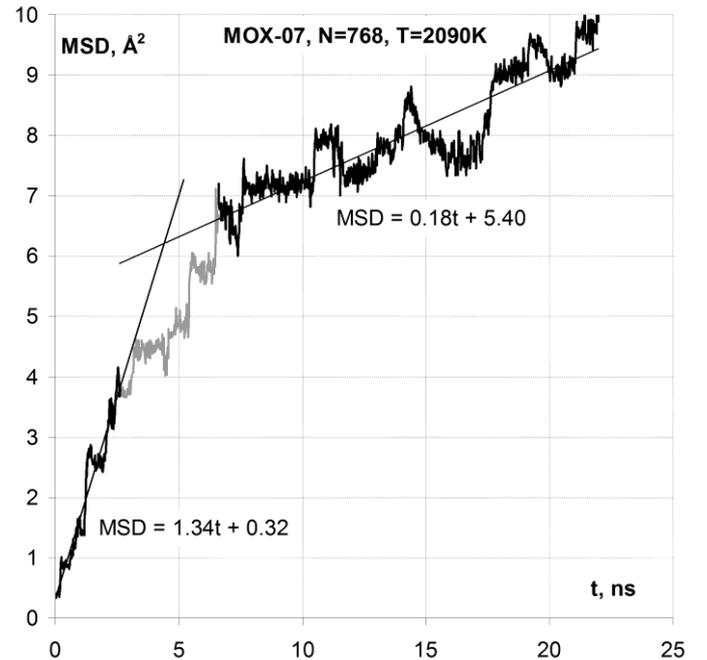

Fig. 13. Time evolution of MSD of surface cations in very small nanocrystal.

Since the observed change in MSD slope could be due to different reasons, in order to verify our assumption we carried out additional simulation, where the particles at vertices, edges and faces of the crystal were delimited by the number of nearest neighbors. Fig. 14 shows that the corresponding curves have different slopes, and the upper curve (corresponding to both vertices and edges) reached saturation just after 3 ns. However, the fraction of particles at vertices and edges is much less for N = 4116, so the MSD curve calculated for all surface particles can not be divided into intervals as easily as in Fig. 13 for N = 768.

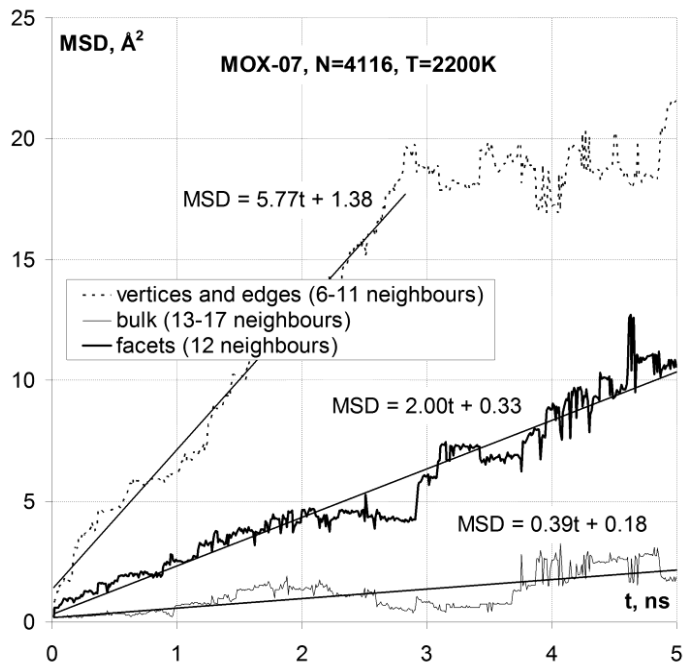

Fig. 14. Time evolution of MSD of cations with different number of neighbors.

Fig. 15 shows that the model dependences are lower than the recommended Matzke at least by an order (which may be due to implicit account of evaporation and condensation of $UO_2$ in the experiments), and the activation energy is lower by approximately 1 eV (see table 6). The location of the experimental data is best reproduced by Goel-08 SPP, and the slope – by MOX-07 SPP. Low activation energy of surface diffusion probably indicates that migration occurs without formation of additional defects. Reducing of the temperature scale to $T_{melt}$ led to convergence of the curves for different SPPs (see Fig. 15b) as in the case of volume cations (Fig. 8).

## 4. Conclusion

In this work simulations were run on a specially designed software package IDGPU, which provides the significant speedup compared with traditional CPU calculations due to the use of high-performance graphics processors of AMD Radeon and NVIDIA GeForce series via the Microsoft DirectCompute parallel computing technology. This allowed for the first time to simulate the intrinsic disordering of cation sublattice in the solid phase of $UO_2$ without the need to create artificial defects. The simulation times reached 2200 ns (440 million MD steps), which allowed to calculate the cation diffusion coefficients down to $4 \times 10^{-11}$ cm$^2$/s with a temperature step of 1–10 K.

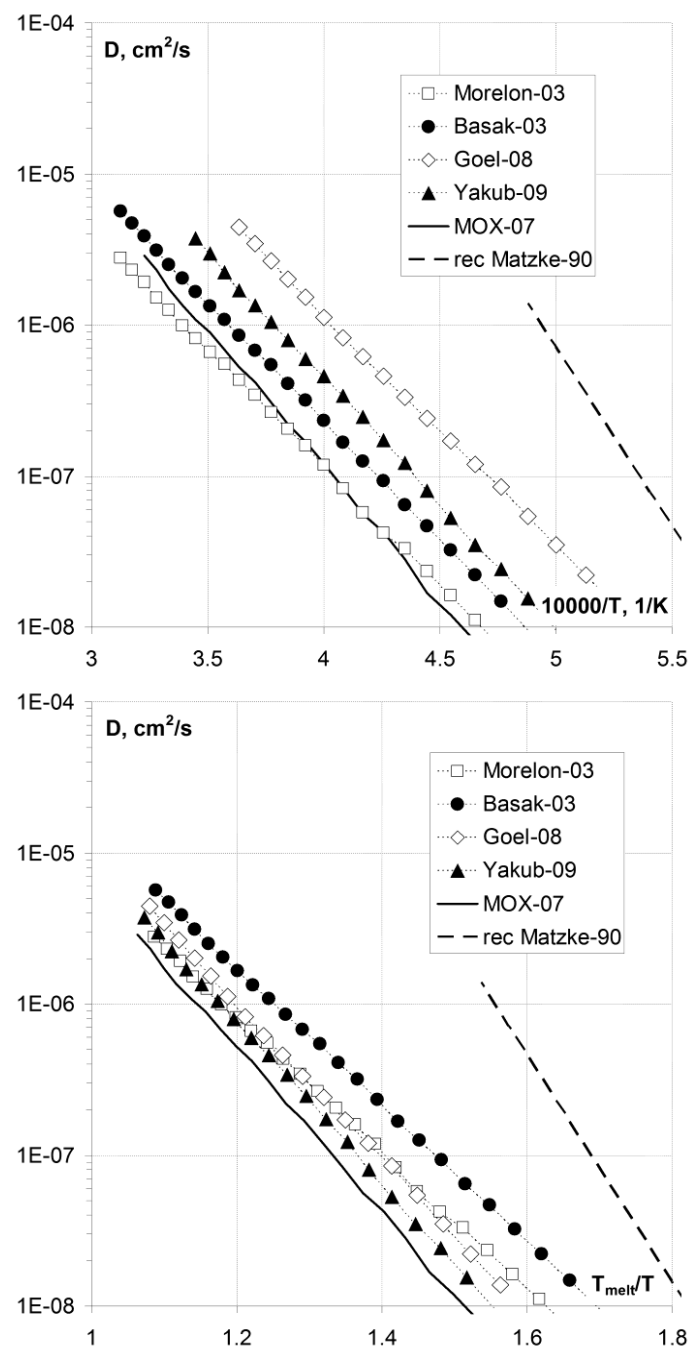

Fig. 15. Temperature dependence of diffusion coefficient of cations at nanocrystal surface. a) traditional temperature scale; b) temperature reduced to $T_{melt}$.

Our computational experiments together with direct visual observations showed that quasi-infinite defectless crystals without surface (simulated under PBC) have no long-lived intrinsic defects. Moreover, regardless of SPP at temperatures up to the melting point the cation diffusion under PBC occurs via the exchange mechanism with formation and recombination of short-lived Frenkel defects, similar to the anion diffusion [4]. The cation diffusion activation energies in the range of 15–22 eV were close to the Frenkel defect formation energies calculated for the corresponding SPP.

The nanocrystals isolated in vacuum (under IBC) allowed modeling the intrinsic Schottky disordering of the crystal lattice with immersing of cation vacancies from the free surface to the bulk. However, besides the vacancy diffusion we found traces of the exchange mechanism near the melting temperature. The effective activation energies of the bulk diffusion were 10.8±1.0 eV and 12.9±1.1 eV for MOX-07 and Yakub-09 SPPs correspondingly.

The non-stoichiometric nanocrystals of $UO_{2.10}$, $UO_{2.15}$ and $UO_{1.85}$ were also simulated. The activation energy of cation diffusion in them was significantly decreased down to 6–7 eV compared to the stoichiometric nanocrystals. Our observations showed that vacancy diffusion is dominant in all of them, and no traces of the cation interstitial mechanism were found (although it was expected to occur in $UO_{1.85}$). The effective number of anion vacancies accompanying the cation vacancy appeared to be two in $UO_{2.00}$ and four in $UO_{2.15}$ and $UO_{1.85}$. Observations at a sufficiently low temperature (necessary to make the anions not as mobile as in the superionic phase) revealed that cation vacancy forms different energetically optimal clusters depending on the stoichiometry: single uranium vacancy in $UO_{2.15}$, uranium vacancy with one oxygen vacancy in $UO_{2.00}$ and uranium vacancy with four oxygen vacancies in $UO_{1.85}$; while the aforementioned values of the effective number of anion vacancies are explained by distortion of the neighbor ions, which screen the electrostatic potential of the point defects.

It is shown that the interstitial mechanism of uranium diffusion in the hypostoichiometric crystals is not necessarily the dominant. The effective energy of formation of the cluster with one cation vacancy and four anion vacancies appeared to be sufficiently low to make the vacancy diffusion dominant in the model nanocrystals of $UO_{1.85}$. It can be an example of cluster mechanisms, which were supposed to occur in $UO_{2-x}$ according to the earlier works [8] [38].

The formulas (2) usually used to calculate the diffusion activation energy in $UO_{2-x}$ appeared to be not applicable to our temperature dependences, because they assume a low concentration of all defects, exponentially increasing with temperature. However, formulas suggested in this work enabled us to describe the results of MD simulation obtained for the superionic phase and predicted a gradual change in the diffusion activation energy in the range of superionic transition. The corresponding extrapolation of our temperature dependence of diffusion coefficient for MOX-07 SPP converged with the experimental data at 60–70% of the melting temperature near the superionic transition. Divergence of experimental dependences of different authors from this extrapolation at different temperatures is probably due to the different concentration of the temperature-independent defects (such as impurities, dislocations or grain boundaries).

Comparison of temperature dependences of the diffusion coefficient obtained in different conditions (under PBC and IBC, with different sets of pair potentials) revealed that they converge to each other when using the reduced temperature scale. On the base of the similar relationship it was concluded [55] that cation diffusion coefficients reach $10^{-9}$–$10^{-8}$ cm$^2$/s at the melting point in the compounds with the fluorite structure (including $UO_2$). Our temperature dependences obtained for the nanocrystals agree with this conclusion, but the extrapolations of the experimental data on $UO_2$ give much lower diffusion coefficients. This fact supports our prediction about increasing of the diffusion activation energy with temperature up to the values close to 10 eV. The possible mechanisms of this increasing are the exchange diffusion and the superionic transition.

The diffusion coefficients of surface cations obtained in this work are lower than the recommended temperature dependence [56], and the activation energies of 3.1–3.6 eV are lower than the recommended value of 4.7 eV.

### *References*

1. H. Mehrer. Diffusion in Solids; Springer: Berlin Heidelberg, 2007.
2. Hj. Matzke. Journal of Chemical Society, Faraday Transactions 2, 83 (1987) 1121.
3. G.E. Murch, C.R.A. Catlow. Journal of Chemical Society, Faraday Transactions 83, 1157 (1987).
4. S.I. Potashnikov, A.S. Boyarchenkov, K.A. Nekrasov, A.Ya. Kupryazhkin. Journal of Nuclear Materials 433 (2013) 215.
5. T. Arima, K. Yoshida, K. Idemitsu, Y. Inagaki, I. Sato. IOP Conference Series: Materials Science and Engineering 9 (2010) 1. doi: 10.1088/1757-899X/9/1/012003.
6. K. Govers, S. Lemehov, M. Hou, M. Verwerft. Journal of Nuclear Materials 395 (2009) 131.
7. E. Yakub, C. Ronchi, D. Staicu. Journal of Chemical Physics 127 (2007) 094508.
8. Hj. Matzke. Journal de physique 34 (1973) 317.
9. Hj. Matzke. Science of Advanced LMFBR fuels, Monograph on Solid State Physics, Chemistry and Technology of Carbides, Nitrides and Carbonitrides of Uranium and Plutonium (North Holland, Amsterdam, 1986).
10. C. Ronchi, M. Sheindlin, D. Staicu, M. Kinoshita. Journal of Nuclear Materials 327 (2004) 58.

11. A.C.S. Sabioni, W.B. Ferraz, F. Millot. Journal of Nuclear Materials 257 (1998) 180.
12. L. Leibowitz, J.K. Fink, O.D. Slagle. Journal of Nuclear Materials 116 (1983) 324.
13. A.C.S. Sabioni, W.B. Ferraz, F. Millot. Journal of Nuclear Materials 278 (2000) 364.
14. P. Sindzingre, M.J. Gillan. Journal of Physics C: Solid State Physics 21 (1988) 4017.
15. K. Yamada, K. Kurosaki, M. Uno, S. Yamanaka. Journal of Alloys and Compounds 307 (2000) 10.
16. K. Yamada, K. Kurosaki, M. Uno, S. Yamanaka. Journal of Alloys and Compounds 307 (2000) 1.
17. S.I. Potashnikov, A.S. Boyarchenkov, K.A. Nekrasov, A.Ya. Kupryazhkin. ISJAEE 5 (2007) 86. http://isjaee.hydrogen.ru/pdf/AEE0507/AEE05-07_Potashnikov.pdf
18. A.Ya. Kupryazhkin, A.N. Zhiganov, D.V. Risovany, K.A. Nekrasov et al. Journal of Nuclear Materials 372 (2008) 233.
19. A.S. Boyarchenkov, S.I. Potashnikov, K.A. Nekrasov, A.Ya. Kupryazhkin. Journal of Nuclear Materials 421 (2012) 1.
20. E. Vincent-Aublant, J-M. Delaye, L. Van Brutzel. Journal of Nuclear Materials 392 (2009) 114.
21. J. Durinck, M. Freyss, P. Garcia. Atomic transport simulations in UO 2+x by ab-initio: oxygen and uranium atomic migration, Tech. Report, Internal Report CEA-SESC/LCC 07-009, 2007.
22. D.A. Andersson, B.P. Uberuaga, P.V. Nerikar. Physical Review B 84 (2011) 054105.
23. P. Contamin, J. Bacmann, J. Marin. Journal of Nuclear Materials 42 (1972) 54.
24. G.L. Reynolds, B. Burton. Journal of Nuclear Materials 82 (1979) 22.
25. T.G. Desai, P. Millett, M. Tonks, D. Wolf. Acta Materialia 58 (2010) 330.
26. S.I. Potashnikov, A.S. Boyarchenkov, K.A. Nekrasov, A.Ya. Kupryazhkin. Journal of Nuclear Materials 419 (2011) 217.
27. A.S. Boyarchenkov, S.I. Potashnikov, K.A. Nekrasov, A.Ya. Kupryazhkin. Journal of Nuclear Materials 427 (2012) 311.
28. P. Goel, N. Choudhury, S.L. Chaplot. Journal of Nuclear Materials 377 (2008) 438.
29. N.-D. Morelon, D. Ghaleb, J.-M. Delhaye, L. Van Brutzel. Philosophical Magazine 83 (2003) 1533.
30. C.B. Basak, A.K. Sengupta, H.S. Kamath. Journal of Alloys and Compounds 360 (2003) 210.
31. S.I. Potashnikov, A.S. Boyarchenkov, K.A. Nekrasov, A.Ya. Kupryazhkin. ISJAEE 8 (2007) 43. http://isjaee.hydrogen.ru/pdf/AEE0807/AEE08-07_Potashnikov.pdf
32. E. Yakub, C. Ronchi, D. Staicu. Journal of Nuclear Materials 389 (2009) 119.
33. A.S. Boyarchenkov, S.I. Potashnikov. Numerical methods and programming 9 (2009) 9. http://num-meth.srcc.msu.ru/english/zhurnal/tom_2009/v10r102.html
34. A.S. Boyarchenkov, S.I. Potashnikov. Numerical methods and programming 10 (2009) 158. http://num-meth.srcc.msu.ru/english/zhurnal/tom_2009/v10r119.html
35. A.S. Boyarchenkov, S.I. Potashnikov. Ionic Dynamics Simulation on Graphics Processing Units. http://code.google.com/p/idgpu/ (last accessed on 28 May 2013).
36. J.R. Matthews, J. Chem. Soc., Faraday Trans. 2 87 (1987) 1273–1285.
37. K. Govers, S. Lemehov, M. Hou, M. Verwerft. Journal of Nuclear Materials 366, 161 (2007).
38. C.R.A. Catlow. Proceedings of the Royal Society of London. Series A, Mathematical and Physical Sciences 353 (1977) 533–561.
39. R.A. Jackson, C.R.A. Catlow, A.D. Murray. J. Chem. Soc., Faraday Trans. 2 83 (1987) 1171–1176.
40. F. Devynck, M. Iannuzzi, M. Krack. Physical Review B 85 (2012) 184103.
41. H. Bracht, S.P. Nicols, W. Walukiewicz, J.P. Silveira et al. Nature (London) 408 (2000) 69.
42. D. Weiler, H. Mehrer. Philosophical Magazine A 49 (1984) 309.
43. A. Chroneos, H. Bracht. Journal of Applied Physics 104 (2008) 093714.
44. P. Nerikar, T. Watanabe, J. S. Tulenko, S. R. Phillpot, S. B. Sinnott. Journal of Nuclear Materials 384 (2009) 61.
45. B. Dorado, J. Durinck, P. Garcia, M. Freyss, M. Bertolus. Journal of Nuclear Materials 400 (2010) 103.
46. T. Petit, C. Lemaignan, F. Jollet, B. Bigot, A. Pasturel. Philosophical Magazine B 77 (1998) 779–786.
47. R.A. Jackson, A.D. Murray, J.H. Harding, C.R.A. Catlow. Philosophical magazine 53 (1986) 27.
48. B. Dorado, D.A. Andersson, C.R. Stanek, M. Bertolus, B.P. Uberuaga, G. Martin, M. Freyss, P. Garcia. Physical Review B 86 (2012) 035110.
49. D. Sheppard, R. Terrel, G. Henkelman. Journal of Chemical Physics 128 (2008) 134106.
50. P. Lovera, C. Ferry, C. Poinssot, L. Johnson. Synthesis report on the relevant diffusion coefficients of fission products and helium in spent nuclear fuels, CEA report CEA-R-6039 (2003) ISSN 0429-3460.
51. M.T. Hutchings, K. Clausen, M.H. Dickens, W. Hayes, J.K. Kjems, P.G. Schnabel, C. Smith. J. Phys. C: Solid State Phys. 17 (1984) 3903–3940.
52. D. Glasser-Leme and Hj. Matzke. Solid State Chem. 3 (1982) 201.
53. D.K. Reimann, T.S. Lundy. J. Am. Ceram. Soc. 52 (1969) 511.
54. Hj. Matzke. Journal de Physique C7-12 (1976) 452.
55. K. Ando, Y. Oishi. Journal of Nuclear Science and Technology 20 (1983) 973–982.
56. Hj. Matzke. J. Chem. Soc. Faraday Trans. 86 (1990) 1243–1256.
57. T. Matsumoto, T. Arima, Y. Inagaki, K. Idemitsu, M. Kato, T. Uchida. Journal of Nuclear Materials (in press). http://dx.doi.org/10.1016/j.jnucmat.2013.04.019
58. T. Arima, S. Yamasaki, Y. Inagaki, K. Idemitsu. Journal of Alloys and Compounds 415 (2006) 43–50.

59. S. Yamasaki, T. Arima, K. Idemitsu, Y. Inagaki. International Journal of Thermophysics 28 (2007) 2.
60. Thermodynamic and Transport Properties of Uranium Dioxide and Related Phases, IAEA (1965). http://books.google.ru/books?id=1vMiAQAAIAAJ
61. M. Baichi, C. Chatillon, G. Ducros, K. Froment. Journal of Nuclear Materials 349 (2006) 57.
62. J.F. Marin, P. Contamin. Journal of Nuclear Materials 30 (1969) 16.
63. J.S. Anderson. Problems in non-stoichiometry (ed. N. Rabenau). North Holland Pub. Co. (1972). 292 p.
64. J. Wang, U. Becker. Journal of Nuclear Materials 433 (2013) 424.
65. S. Yu. Davydov. Physics of the solid state 41 (1999) 8.

Table 1. Point defects formation energies (in eV), calculated by the lattice statics method in comparison with cation diffusion activation energies obtained by MD simulations under PBC.

| Pair potentials | Schottky | Frenkel | Anti-Frenkel | Activation energy (MD) |
|---|---|---|---|---|
| Walker-81 | 7.8 (8.3) | 22.1 (22.3) | 5.9 (6.0) | – |
| Busker-02 | 14.7 | 29.0 | 8.4 | – |
| Nekrasov-08 | 8.7 | 21.0 | 5.7 | – |
| Morelon-03 | 8.0 (8.0) | 15.6 (15.7) | 3.9 (3.9) | 17.9 |
| Yamada-00 | 12.9 (13.5) | 18.3 (18.5) | 5.8 (6.0) | – |
| Basak-03 | 10.3 (10.8) | 16.8 (17.0) | 5.8 (6.0) | 16.7 |
| Arima-05 | 14.5 | 23.0 | 7.9 | – |
| Goel-08 | 7.7 | 17.6 | 5.2 | 21.9 |
| Yakub-09 | 10.9 | 15.9 | 5.6 | 15.3 |
| MOX-07 | 9.8 | 15.6 | 4.1 | 15.1 |
| rec Matzke-87 [2] | 6.5±0.5 | 9.5±0.5 | 3.5±0.5 | – |
| rec Matthews-87 [36] | 5±1, 7±1 | 10±1 | 3.8 | – |
| DFT Nerikar-09 [44] | 7.6 | 15.1 | 3.95 | – |
| DFT Dorado-10 [45] | 10.6 | 14.6 | 6.5 | – |

( ) – values in parentheses are cited from [37] for comparison;
rec – recommendations based on experimental data;
DFT – results of *ab initio* calculations using the Density Functional Theory.

Table 2. Energies of interstitial and vacancy migration of cations (in eV), calculated by the lattice statics and molecular dynamics under PBC.

| Pair potentials | Lattice statics | | Molecular dynamics | |
|---|---|---|---|---|
| | interstitial | vacancy | interstitial | vacancy |
| Morelon-03 | 4.2 (4.2) | 3.9 (3.9) | 1.4 | 3.7 |
| Basak-03 | 5.6 (5.1) | 4.2 (5.7) | 1.2 | 2.9 |
| Goel-08 | 4.2 | 4.1 | 0.9 | 3.8 |
| Yakub-09 | 5.7 | 3.7 | 1.7 | 3.1 |
| MOX-07 | 4.5 | 3.7 | 1.0 | 2.9 |
| rec Matzke-87 [2] | low | 2.4 | – | |
| rec Matthews-87 [36] | 0.5–2.0 | 1.5–2.5 | | |
| DFT Dorado-12 [48] | 4.4 | 4.2 | | |
| DFT Dorado-10 [45] | 5.8 | 4.4 | | |

( ) – values in parentheses are cited from [37] for comparison;
rec – recommendations based on experimental data;
DFT – results of *ab initio* calculations using the Density Functional Theory.

Table 3. Cation diffusion activation energies (in eV), calculated within the Lidiard-Matzke model in comparison with experimental data.

| Pair potentials | Vacancy mechanism | | | Interstitial mechanism |
|---|---|---|---|---|
| | $UO_{2+x}$ | $UO_2$ | $UO_{2-x}$ | $UO_{2-x}$ |
| Morelon-03 | 3.9 | 7.8 | 11.7 | 9.0 |
| Basak-03 | 1.6 | 7.4 | 13.2 | 7.7 |
| Goel-08 | 1.1 | 6.3 | 11.5 | 10.8 |
| Yakub-09 | 2.8 | 8.4 | 14.0 | 6.7 |
| MOX-07 | 4.5 | 8.6 | 12.7 | 6.8 |
| exp [62] [50] [53] [11] | 3.5–5.0 | 3.7–5.0 | – | – |
| exp Matzke-87 [2] | 2.6 | 5.6 | 7.8 | 5.0 |
| calc Matzke-87 [2] | 1.9 | 5.4 | 8.9 | 4.3 |
| calc Matthews-87 [36] | 1.4 | 5.2 | 9.0 | 4.3 |
| calc Dorado-12 [48] | 3.6 | 6.9 | 10.2 | 9.6 |
| calc Dorado-10 [45] | 2.0 | 8.5 | 15.0 | 9.8 |

exp – experimental data;
calc – calculation based on the values of formation and migration energies from the corresponding work.

Table 4. Cation self-diffusion characteristics in superionic phase of $UO_2$, simulated under PBC.

| Pair potentials | T, K | $E_D$, eV | $D_0$, cm²/s |
|---|---|---|---|
| Morelon-03 | 4000–4270 | 17.9±0.8 | $(2.39^{+17.8}_{-2.11}) \times 10^{13}$ |
| Basak-03 | 3950–4200 | 16.7±0.7 | $(5.40^{+30.3}_{-4.58}) \times 10^{12}$ |
| Goel-08 | 3650–3860 | 21.9±1.6 | $(1.59^{+226}_{-1.58}) \times 10^{20}$ |
| Yakub-09 | 3450–3740 | 15.3±0.5 | $(1.64^{+6.84}_{-1.32}) \times 10^{13}$ |
| MOX-07 | 3560–4010 | 15.1±0.3 | $(4.01^{+6.83}_{-2.53}) \times 10^{11}$ |

Table 5. Cation self-diffusion characteristics in melted $UO_2$, simulated under PBC.

| Pair potentials | T, K | $E_D$, eV | $D_0$, cm²/s |
|---|---|---|---|
| Morelon-03 | 4250–5250 | 1.07±0.01 | $(1.21^{+0.04}_{-0.04}) \times 10^{-3}$ |
| Basak-03 | 4200–5200 | 1.20±0.01 | $(1.59^{+0.05}_{-0.05}) \times 10^{-3}$ |
| Goel-08 | 3850–4850 | 0.98±0.01 | $(1.22^{+0.03}_{-0.03}) \times 10^{-3}$ |
| Yakub-09 | 3750–4750 | 1.12±0.01 | $(1.77^{+0.05}_{-0.05}) \times 10^{-3}$ |
| MOX-07 | 4000–5000 | 1.11±0.01 | $(1.42^{+0.04}_{-0.04}) \times 10^{-3}$ |

Table 6. Self-diffusion characteristics of cations at surface of $UO_2$ nanocrystals.

| Pair potentials | T, K | $E_D$, eV | $D_0$, cm²/s |
|---|---|---|---|
| Morelon-03 | 2200–3200 | 3.12±0.07 | $(2.27^{+0.83}_{-0.61}) \times 10^{-1}$ |
| Basak-03 | 2200–3180 | 3.13±0.05 | $(4.80^{+1.20}_{-0.96}) \times 10^{-1}$ |
| Goel-08 | 1960–2680 | 3.07±0.04 | $(1.91^{+0.37}_{-0.31}) \times 10^{0}$ |
| Yakub-09 | 2020–2870 | 3.31±0.06 | $(2.12^{+0.66}_{-0.50}) \times 10^{0}$ |
| MOX-07 | 2100–3100 | 3.56±0.04 | $(1.77^{+0.30}_{-0.26}) \times 10^{0}$ |